\documentclass[lettersize,journal]{IEEEtran}
\usepackage{amsmath,amsfonts}
\usepackage{algorithmic}
\usepackage{array}
\usepackage[caption=false,font=normalsize,labelfont=sf,textfont=sf]{subfig}
\usepackage{textcomp}
\usepackage{stfloats}
\usepackage{url}
\usepackage{verbatim}
\usepackage{graphicx}
\usepackage{booktabs}
\def\BibTeX{{\rm B\kern-.05em{\sc i\kern-.025em b}\kern-.08em
    T\kern-.1667em\lower.7ex\hbox{E}\kern-.125emX}}
\usepackage{balance}

\begin{document}
\title{Efficient Multi-Pair IoT Communication with Holographically Enhanced Meta-Surfaces Leveraging OAM Beams: Bridging Theory and Prototype}
\author{Yufei Zhao,
        Yong Liang Guan,~\IEEEmembership{Senior Member,~IEEE,}
        Afkar Mohamed Ismail,
        Gaohua Ju,
        Deyu Lin,~\IEEEmembership{Member,~IEEE,}
        Yilong Lu,~\IEEEmembership{Fellow,~IEEE,}
        and~Chau Yuen,~\IEEEmembership{Fellow,~IEEE}
\thanks{This work has been submitted to IEEE Internet of Things Journal, 2023. (Corresponding author: Yufei Zhao, e-mail: yufei.zhao@ntu.edu.sg).}
\thanks{Yufei Zhao, Yong Liang Guan, Afkar Mohamed Ismail, Yilong Lu, and Chau Yuen are with School of Electrical and Electronic Engineering, Nanyang Technological University, 639798, Singapore.

Gaohua Ju is with the Transport Research Centre, Nanyang Technological University, 639798, Singapore.

Deyu Lin is with School of Electronic Information and Electrical Engineering, Shanghai Jiao Tong University, China, School of Software, Nanchang University, China}
\thanks{Manuscript received 21th, Jul., 2023; revised 19th, Oct. 2023.}}



\maketitle

\begin{abstract}
Meta-surfaces, also known as Reconfigurable Intelligent Surfaces (RIS), have emerged as a cost-effective, low power consumption, and flexible solution for enabling multiple applications in Internet of Things (IoT). However, in the context of meta-surface-assisted multi-pair IoT communications, significant interference issues often arise amount multiple channels. This issue is particularly pronounced in scenarios characterized by Line-of-Sight (LoS) conditions, where the channels exhibit low rank due to the significant correlation in propagation paths. These challenges pose a considerable threat to the quality of communication when multiplexing data streams.
In this paper, we introduce a meta-surface-aided communication scheme for multi-pair interactions in IoT environments. Inspired by holographic technology, a novel compensation method on the whole meta-surface has been proposed, which allows for independent multi-pair direct data streams transmission with low interference. To further reduce correlation under LoS channel conditions, we propose a vortex beam-based solution that leverages the low correlation property between distinct topological modes. We use different vortex beams to carry distinct data streams, thereby enabling distinct receivers to capture their intended signal with low interference, aided by holographic meta-surfaces. Moreover, a prototype has been performed successfully to demonstrate two-pair multi-node communication scenario operating at 10 GHz with QPSK/16-QAM modulation. The experiment results demonstrate that, even under LoS conditions, the isolation between the two-pair channels exceeds 21 dB. This allows receiving users to undertake simultaneous, same-frequency multiplexed data transmission under extremely low interference conditions, with a real-time demodulation Bit Error Rate (BER) remaining below $3.8 \times 10^{-3}$ at achievable Signal-to-Noise Ratio (SNR) conditions. Through the convergence of holographic meta-surfaces and vortex beams, we present a fresh perspective on achieving efficient, low-interference multi-pair IoT communications.
\end{abstract}


\section{Introduction}
\begin{figure}[htbp]
\centering
\includegraphics[width=3.5in]{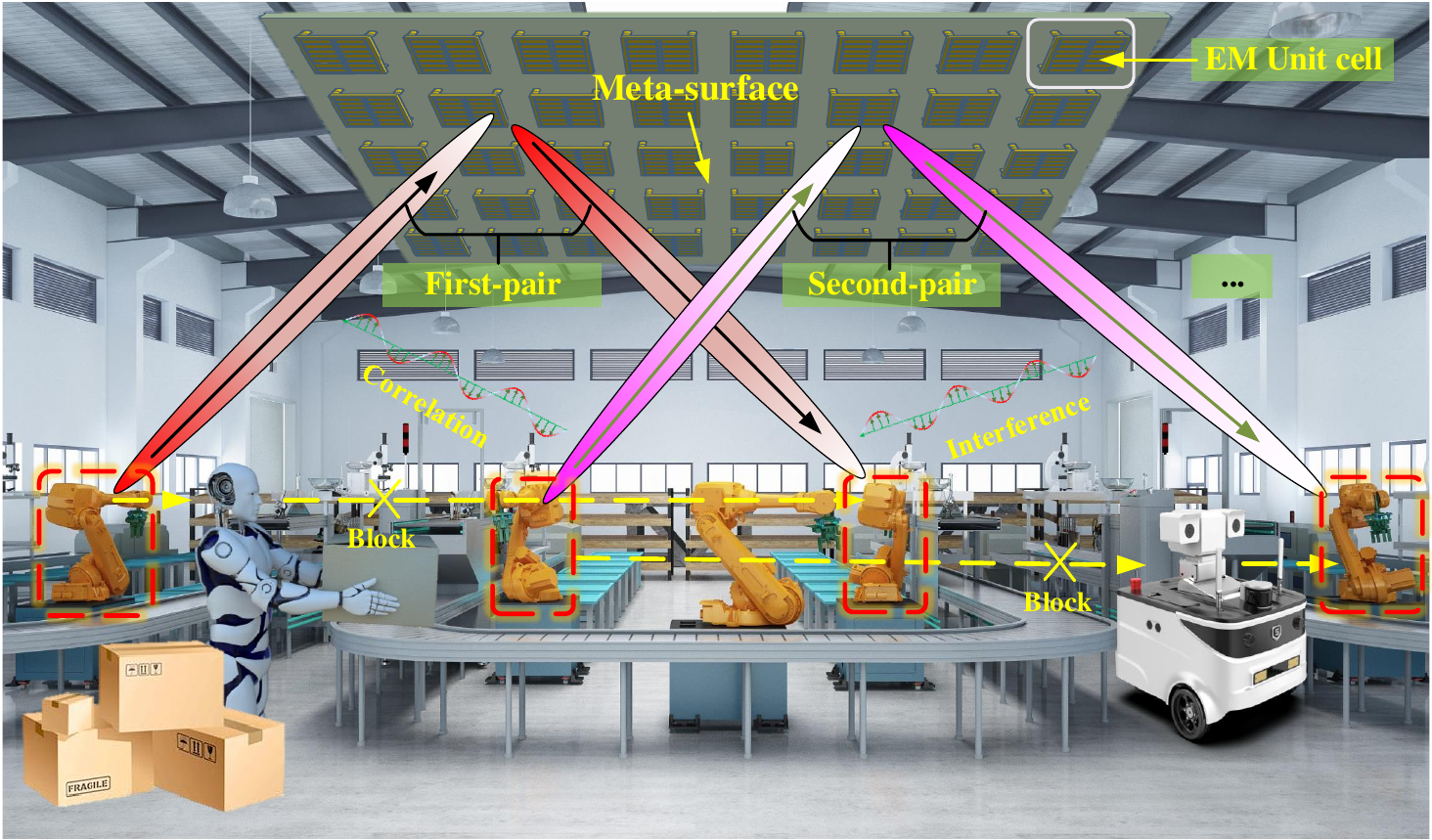}
\caption{Mate-surface-aided multi-pair wireless communications in IoT.}
\label{fig1}
\end{figure}

\IEEEPARstart{T}{he} Internet of Things (IoT), a rapidly expanding field that has experienced significant growth in recent years, is reshaping numerous sectors, including healthcare, transportation, smart homes, and industrial automation, etc. Within IoT systems, there are numerous intelligent nodes, which typically rely on wireless connections such as mobile cellular networks, Bluetooth, Wi-Fi, and Zigbee to exchange data, forming intricate and complex communication networks \cite{IoT}. Factory workshops, supermarkets, logistics warehouses, transportation hubs, and more are all densely populated with IoT technologies. Within these environments, challenges emerge due to the intricacies of the physical surroundings. Factors such as physical obstacles, signal interference can impede the seamless transmission of data. It is in these complex scenarios that the role of the meta-surface becomes pivotal. As illustrated in Fig. \ref{fig1}, when Line of Sight (LoS) communication links between IoT nodes are obstructed, meta-surfaces have the capacity to adapt the traditional physical propagation environment as needed, enabling the flexible establishment of additional LoS communication paths \cite{Cui1,Yuen}.

Meta-surfaces, also named as Reconfigurable Intelligent Surfaces (RIS), offer several benefits including low cost, low power consumption, ease of deployment, and reconfigurability. With the assistance of meta-surfaces, low-power IoT nodes can extend signal coverage, mitigate interference, and facilitate extensive Device-to-Device (D2D) communications \cite{Cui1,Yuen,Wu,IoT2}. Generally, meta-surfaces consist of a two-dimensional array of sub-wavelength elements capable of manipulating the phase and amplitude of Electro-Magnetic (EM) waves with high accuracy and flexibility. This capability allows meta-surfaces to programmatically reflect, refract, or focus EM waves, making them suitable for numerous IoT applications such as wireless power transfer \cite{power}, localization, sensing \cite{Huang}, and wideband communications \cite{THz}. Despite these advantages, using meta-surfaces in multi-pair wireless communications services poses unique challenges, particularly under LoS channel conditions.

To ensure the stability of meta-surface-aided communication links and facilitate precise control of the meta-surface's status by the transmitter, the meta-surface is typically deployed within the visible range of the transmitter. In other words, meta-surface-aided communication links are characteristic of LoS channels, or direct paths \cite{los}. Clearly, LoS channels lack scattering components, exhibit low-rank characteristics, and possess strong correlations among multiple data streams. In IoT application scenarios, it is evident that the inherent low-rank nature of LoS channels imposes significant constraints on the diversity and multiplexing capabilities of multi-pair serving scenarios. In environments with limited scattering, the LoS channel responses tend to exhibit strong correlation, thereby diminishing the degrees of freedom available to multiple antenna systems \cite{jinshi,backhaul}. The high correlation of the LoS channel means that the signals from different Tx nodes are highly correlated when arriving at the meta-surface, which causes severe interference for different users serviced by the same meta-surface. Considering Shannon's information theory, the inherent low-rank property of the LoS channel implies a limitation in the number of independent propagation paths, which further reduces the total capacity of the whole multi-user communication system.

The existing body of literature has proposed several solutions to address the challenges associated with meta-surface-aided systems under LoS conditions. For example, cooperative systems like the double-RIS aided communications have been introduced to increase data rates \cite{double-IRS1,double-IRS2}. In \cite{PIN}, researchers propose the antennas/PINs selection scheme and design the joint beamforming method to increase the spectrum efficiency under LoS conditions. However, these solutions demand more resources for pilot signals and control signal overhead to ensure effective cooperation and synchronization among the different RISs. Furthermore, they do not adequately address the problem of high correlation in the LoS channel, which still leads to severe interference for different users serviced by the same meta-surface. This constraint on multiplexing capabilities of multi-pair serving systems negatively impacts the total capacity of the whole communication system.
Some researchers anticipate that increasing the electrical size of the meta-surface can help reduce the spatial correlation between Tx nodes and the meta-surface, and predict the application and performance of large intelligent surfaces in the future \cite{maxiaoli,zengyong2}. It is undeniable that the large surface offers superior angular resolution in the spatial domain, thereby reducing the channel correlation between multiple users, but it also obviously increases the system overhead and control complexity.

In recent years, researchers have been extensively drawn to the low-complexity and high-capacity transmission capabilities of vortex beams, especially under LoS conditions \cite{chao1,NTT}. The vortex beam, also known as Orbital Angular Momentum (OAM), is considered to be an inherent property of EM waves \cite{cheng1}. By exploiting the low correlation property between different topology modes, vortex beams carrying different data streams can serve distinct receivers with lower interference \cite{chen1,zhao1}. In this paper, we propose a meta-surface-aided communication scheme that utilizes vortex beams to transmit data under the LoS channel conditions between multiple Tx nodes and the meta-surface.

Furthermore, it is reasonable to assume that all the Tx nodes, Rx users, and meta-surface possess the angle of arrival (AoA) and angle of departure (AoD) information associated with their respective LoS paths, given that their relative positions remain constant. Then, for the multi-pair nodes communications, addressing each node's unique AoA, AoD and vortex mode requires tailored phase compensation, posing a significant challenge when attempting to perform simultaneous phase compensation for all transmitting nodes using a single meta-surface. The meta-surface is confronted with the formidable task of directing the phase-compensated signals towards distinct users, a necessity for supporting multi-user service requirements within the IoT system. This demand adds another layer of complexity to both the hardware and software design of the meta-surface.

To solve this trick problem, we put forth a meta-surface design methodology that draws inspiration from holographic principles. The meta-surface's compensation process can be envisioned as a hologram brought to life by the interplay between the 'reference wave' emanating from the Tx node, and the 'object wave' produced by the Rx user at the meta-surface location. By illuminating the meta-surface with the reference waves, we effectively 'reconstruct' the corresponding object waves, thereby establishing efficient one-to-one mapping communication links that bridge the Tx nodes, the holographic meta-surface, and finally, the Rx users. Simulations and experiments demonstrate that this approach can effectively facilitate low-interference communications between multi-pair Tx-Rx links.

For clarify, the main contributions of this paper are summarized as follows.
\begin{itemize}
\item First, we specifically focused on and explored the common issue of low-rank LoS channels in meta-surface-aided wireless communications. Drawing on EM field theory, we established a channel model for meta-surface-aided multi-pair IoT scenarios. In this model, the EM radiation characteristics of the Tx nodes were thoroughly explored and scrutinized. It was initially proposed that, with the transmit power fixed and the relative positions of Tx nodes, meta-surface, and Rx nodes determined, the radiation patterns of the Tx nodes have a substantial impact on the correlation between wireless channels.
\item Next, we introduce a novel approach that harnesses the orthogonality between vortex beams to reduce correlation between LoS channels for the first time. Our proposal involves adopting microstrip vortex beam antennas at the Tx nodes, replacing the conventional plane wave antennas. This strategy leverages the additional phase variation introduced by vortex wavefront to counteract the similarities between LoS channel direct paths, resulting in a linear increase in capacity as the number of Tx-Rx pairs grows. In practical scenarios, even when the orthogonality between vortex modes is compromised due to physical constraints, using different vortex beam antennas at different Tx nodes substantially decreases the spatial channel correlation among them.
\item Moreover, we introduce a low-complexity compensation method for meta-surfaces in multi-pair IoT service scenarios. Numerical simulations and experimental results indicate that this approach establishes efficient one-to-one mapping communication links, seamlessly connecting the Tx nodes, the holographic meta-surface, and, ultimately, the Rx users. This innovative method has the potential to address the challenges presented by multi-pair service scenarios, ensuring robust and efficient communication links without the need for complex iterative computational processes.
\item Finally, from theory to practice, we also demonstrate the potential of this approach by providing a prototype design for a two-pair communication scenario at 10 GHz with QPSK/16-QAM modulation. Our results showcase over 21 dB isolation between the two-pair IoT nodes channels with the assistance of a meta-surface and 2 distinct vortex modes. The demodulation constellation diagrams and Bit Error Rate (BER) measurement results verify the direct independent data communications between multiple nodes with low interference. Hence, this holograph-inspired meta-surface system with vortex beams has the potential to provide efficient communications services to smart home, automated factory, supermarket, and could become a new type of low-complexity solution for multiple IoT applications.
\end{itemize}

The rest of this paper is organized as follows. Section II presents the system model and describes the meta-surface-aided communication process in the IoT scenarios. Section III introduces the vortex beam generation method and analyzes the theoretical channel capacity with/without vortex beams. Section IV proposes a novel meta-surface design methodology inspired by the holographic technology for multi-pair IoT services. Section V presents the experimental setup and measurement results of the whole demonstration prototype. Finally, Section VI concludes this paper, discusses the limitations and future research directions.

\section{System Architecture and Channel Model}
We consider a typical meta-surface-aided multi-pair wireless communication system, in which the meta-surface acts as the reflection one. Assume that there is no direct path between the Tx nodes and the Rx users due to obstructions such as walls or large furniture blocking the path. At this stage, the meta-surface can be positioned between the Tx node and the Rx user. Owing to its lightweight and low power consumption, the meta-surface can be easily mounted on walls, ceilings, or other simple supports as per convenience. Drawing on numerous established research precedents \cite{yuen2,jinshi1,gaofeifei1}, we make the assumption that the signals propagate from the Tx nodes to the meta-surface through LoS channels, being reflected by the meta-surface, and then transmit towards the Rx users through other distinct LoS channels. In this way, the new communication paths have been established between the Tx nodes and the blocked Rx users.

\begin{figure}[!t]
\centering
\includegraphics[width=3.2in]{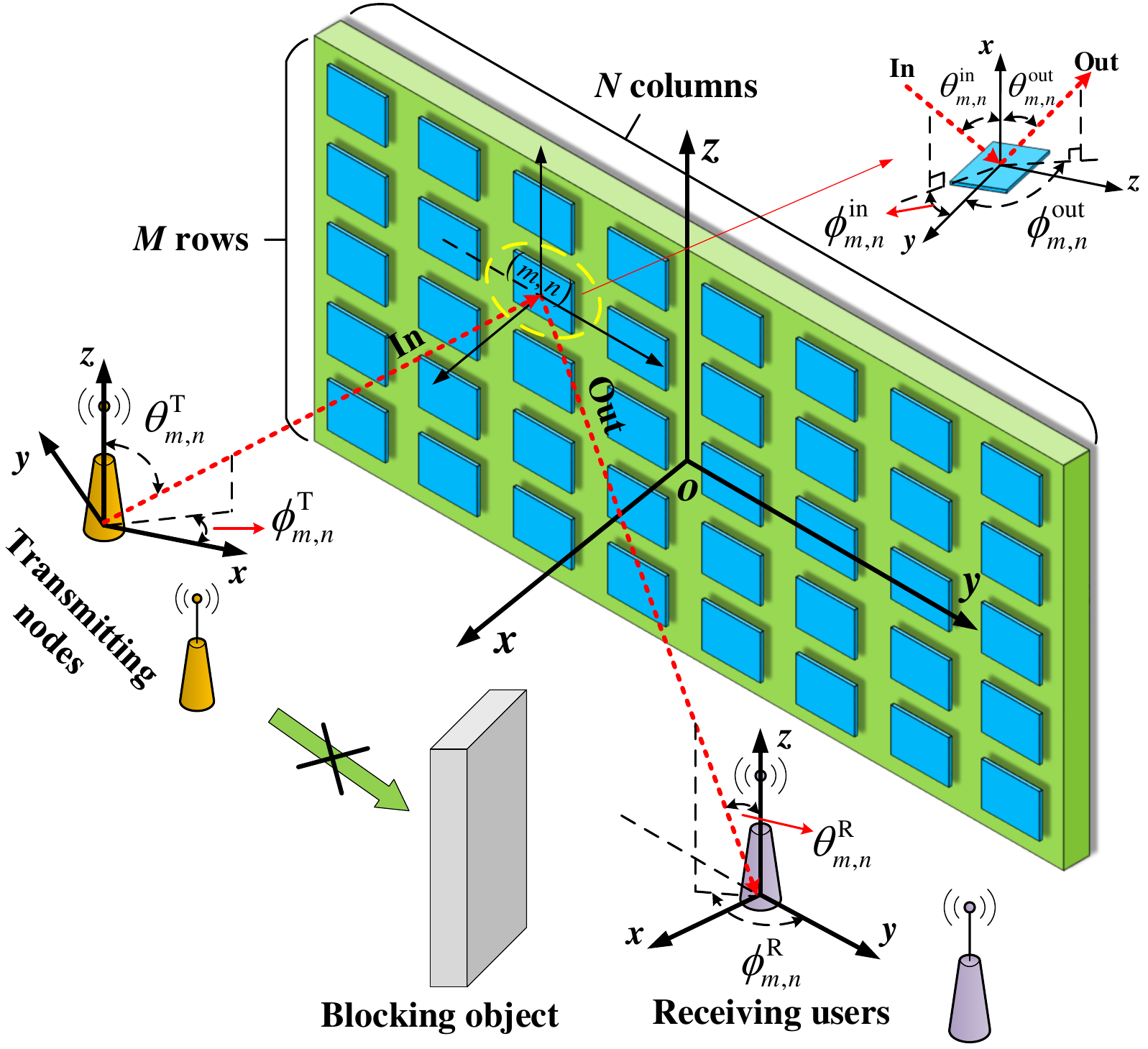}
\caption{System geometrical model of the meta-surface-aided wireless communications scheme.}
\label{fig2}
\end{figure}

As shown in Fig. \ref{fig2}, it is assumed that there are ${N_{\text{T}}}$ Tx nodes in the IoT system, and each node can generate vortex beams carrying different OAM modes, which will be described in detail in Sect. III; there are ${N_{\text{R}}}$ Rx users, and each user is equipped with a directional single antenna. Suppose there is a reflection meta-surface comprising $M \times N$ passive RF elements, with each element having a reflection coefficient ${{\Gamma _{m,n}} = {A_{m,n}}{e^{j{\varphi _{m,n}}}}}$ for the $m$-th row and $n$-th column. The amplitude ${A_{m,n}}$ is fixed and the phase ${{\varphi _{m,n}}}$ can be adjusted independently. In Cartesian coordinate system, assuming that the meta-surface is placed on the $y$-$z$ plane, and the center of the meta-surface is defined as the coordinate origin, the size of each element is ${d_y} \times {d_z}$, then the coordinate of the $m$-th row $n$-th column element can be calculated as
\begin{equation} \label{eq1}
\begin{array}{*{20}{l}}
{{{\bf{u}}_{m,n}} = ({x_{m,n}},{y_{m,n}},{z_{m,n}})} \hfill \\
{\kern 1pt} {\kern 1pt} {\kern 1pt} {\kern 1pt} {\kern 1pt} {\kern 1pt} {\kern 1pt} {\kern 1pt} {\kern 1pt} {\kern 1pt} {\kern 1pt} {\kern 1pt}{\kern 1pt}{\kern 1pt} {\kern 1pt} {\kern 1pt} {\kern 1pt} {\kern 1pt} {\kern 1pt} {\kern 1pt} {\kern 1pt} {\kern 1pt} {\kern 1pt} {\kern 1pt} { = \left[ {0,(m - \frac{{M{\rm{  +  }}1}}{2}){d_y},(n - \frac{{N{\rm{  +  }}1}}{2}){d_z}} \right]}.
\end{array}
\end{equation}

Similarly, the coordinates of the Tx nodes and the Rx users can be denoted as
\begin{align} \label{eq2-3}
  {{\mathbf{u}}_{{\text{T}},{n_{\text{T}}}}} &= ({x_{{n_{\text{T}}}}},{y_{{n_{\text{T}}}}},{z_{{n_{\text{T}}}}}),{\kern 1pt} {\kern 1pt} {\kern 1pt} {\kern 1pt} {n_{\text{T}}} \in \left\{ {1,2, \ldots ,{N_{\text{T}}}} \right\}, \hfill \\
  {{\mathbf{u}}_{{\text{R}},{n_{\text{R}}}}} &= ({x_{{n_{\text{R}}}}},{y_{{n_{\text{R}}}}},{z_{{n_{\text{R}}}}}),{\kern 1pt} {\kern 1pt} {\kern 1pt} {\kern 1pt} {n_{\text{R}}} \in \left\{ {1,2, \ldots ,{N_{\text{R}}}} \right\}. \hfill
\end{align}
Hence, for each unit element on the meta-surface, the distance ${d_{{n_{\text{T}}},m,n}}$ from the $n_{\text{T}}$-th Tx note, and the distance ${d_{m,n,{n_{\text{R}}}}}$ to the $n_{\text{T}}$-th Rx users can be simply calculated
\begin{align} \label{eq4-5}
\begin{tiny}
\begin{gathered}
  {d_{{n_{\text{T}}},m,n}} = \sqrt {{{\left( {{x_{{n_{\text{T}}}}}} \right)}^2} + {{\left( {{y_{{n_{\text{T}}}}} - (m - \frac{{M{\text{  +  }}1}}{2}){d_y}} \right)}^2} + {{\left( {{z_{{n_{\text{T}}}}} - (n - \frac{{N{\text{  +  }}1}}{2}){d_z}} \right)}^2}},  \hfill \\
  {d_{m,n,{n_{\text{R}}}}} = \sqrt {{{\left( {{x_{{n_{\text{R}}}}}} \right)}^2} + {{\left( {{y_{{n_{\text{R}}}}} - (m - \frac{{M{\text{  +  }}1}}{2}){d_y}} \right)}^2} + {{\left( {{z_{{n_{\text{R}}}}} - (n - \frac{{N{\text{  +  }}1}}{2}){d_z}} \right)}^2}}.  \hfill \\
\end{gathered}
\end{tiny}
\end{align}
To simplify the analysis, as depicted in Fig. \ref{fig2}, we assume that both the Tx notes and Rx users radiate towards the center of the meta-surface in the spherical coordinate system. The incident EM wave is characterized by elevation angle $\theta _{m,n}^{{\text{in}}}$ and azimuth angle $\phi _{m,n}^{{\text{in}}}$ with respect to the unit element ${U_{m,n}}$, while the outgoing wave is represented by elevation angle $\theta _{m,n}^{{\text{out}}}$ and azimuth angle $\phi _{m,n}^{{\text{out}}}$. At the Tx node ${T_{{n_{\text{T}}}}}$, the radiation direction pointing towards the unit ${U_{m,n}}$ is denoted by elevation angle $\theta _{m,n}^{{\text{T}}}$ and azimuth angle $\phi _{m,n}^{{\text{T}}}$, and at the Rx user ${R_{{n_{\text{R}}}}}$, $\theta _{m,n}^{{\text{R}}}$ and $\phi _{m,n}^{{\text{R}}}$ denote the elevation and azimuth angles of the receiving direction from the unit ${U_{m,n}}$, respectively. Then, the power of the incident wave from the node ${T_{{n_{\text{T}}}}}$ to the element ${U_{m,n}}$ can be denoted as
\begin{equation} \label{eq6}
\begin{footnotesize}
\begin{gathered}
p_{m,n}^{{n_{\text{T}}}} = \frac{{{g_{{n_{\text{T}}}}}{p_{{n_{\text{T}}}}}}}{{4\pi {{\left( {{d_{{n_{\text{T}}},m,n}}} \right)}^2}}}{F_{{n_{\text{T}}}}}\left( {\theta _{m,n}^{\text{T}},\phi _{m,n}^{\text{T}}} \right){F_{m,n}}\left( {\theta _{m,n}^{{\text{in}}},\phi _{m,n}^{{\text{in}}}} \right){d_y}{d_z},
\end{gathered}
\end{footnotesize}
\end{equation}
where, ${{g_{{n_{\text{T}}}}}}$ and ${{p_{{n_{\text{T}}}}}}$ denote the gain and radiation power of the ${n_{\text{T}}}$-th Tx node, ${F_{{n_{\text{T}}}}}\left( {\theta _{m,n}^{\text{T}},\phi _{m,n}^{\text{T}}} \right)$ and ${F_{m,n}}\left( {\theta _{m,n}^{{\text{in}}},\phi _{m,n}^{{\text{in}}}} \right)$ express the normalized power radiation pattern of ${T_{{n_{\text{T}}}}}$ and ${U_{m,n}}$, respectively. Supposing that the polarizations are always matched, according to (\ref{eq6}), the electric filed on the unit ${U_{m,n}}$ can be calculated as
\begin{equation} \label{eq7}
\begin{footnotesize}
\begin{gathered}
E_{m,n}^{{n_{\text{T}}}} = \sqrt {\frac{{2{\kappa _0}p_{m,n}^{{n_{\text{T}}}}}}{{{d_y}{d_z}}}} {e^{ - jk{d_{{n_{\text{T}}},m,n}}}},
\end{gathered}
\end{footnotesize}
\end{equation}
where, ${{\kappa _0}}$ is the air characteristic impedance, $k = 2\pi /\lambda $ is the wave vector, $\lambda $ is the wave length. As we know, each RF element enjoys its own independently controllable reflection coefficient ${\Gamma _{m,n}}$, and it is the key feature that makes meta-surfaces highly flexible in the field of EM waves modulation. Based on energy conservation principle, the power of the outgoing signal equals to the power of the incident signal multiplied by the square of the modulus of the reflection coefficient, i.e.,
\begin{equation} \label{eq8}
p_{m,n}^{{\text{out}}} = p_{m,n}^{{n_{\text{T}}}} \times {\left| {{\Gamma _{m,n}}} \right|^2}.
\end{equation}
In accordance with waveguide transmission line theory, ${\Gamma _{m,n}}$ can also be regarded as equivalent to S11 or S21 parameters. In other words, the reconfigurable properties of the meta-surface are due to the unique element structure design that enables the alteration of the amplitude or phase of the incident EM waves \cite{Cui1}. We will provide a more detailed explanation and a demo system about this concept in Sect. V.

Furthermore, after the meta-surface, the receiving power of the EM wave from the the element ${U_{m,n}}$ to the Rx user ${R_{{n_{\text{R}}}}}$ can also be denoted as
\begin{equation} \label{eq9}
\begin{footnotesize}
\begin{gathered}
p_{m,n}^{{n_{\text{R}}}} = \frac{{{g_{m,n}}p_{m,n}^{{\text{out}}}}}{{4\pi {{\left( {{d_{{n_{\text{R}}},m,n}}} \right)}^2}}}{F_{m,n}}\left( {\theta _{m,n}^{{\text{out}}},\phi _{m,n}^{{\text{out}}}} \right){F_{{n_{\text{R}}}}}\left( {\theta _{m,n}^{\text{R}},\phi _{m,n}^{\text{R}}} \right)A_{{n_{\text{R}}}}^e.
\end{gathered}
\end{footnotesize}
\end{equation}
Similarly, ${{g_{m,n}}}$ denotes the gain of the element ${U_{m,n}}$, ${F_{{n_{\text{R}}}}}\left( {\theta _{m,n}^{\text{R}},\phi _{m,n}^{\text{R}}} \right)$ is the normalized power radiation pattern of the ${R_{{n_{\text{R}}}}}$ user. Specifically, $A_{{n_{\text{R}}}}^e$ denotes the effective aperture of each independent receiving user ${R_{{n_{\text{R}}}}}$, which can be further expressed as \cite{jinshi}
\begin{equation} \label{eq10}
A_{{n_{\text{R}}}}^e = \frac{{{g_{{n_{\text{R}}}}}{\lambda ^2}}}{{4\pi }}.
\end{equation}
Then, according to (\ref{eq6}), (\ref{eq7}), (\ref{eq8}), (\ref{eq9}), and (\ref{eq10}), the electric field on the Rx user ${R_{{n_{\text{R}}}}}$ after modulated by one single element ${U_{m,n}}$ can be calculated as
\begin{equation} \label{eq11}
\begin{footnotesize}
\begin{gathered}
  E_{m,n}^{{n_{\text{R}}}} = \sqrt {\frac{{2{\kappa _0}p_{m,n}^{{n_{\text{R}}}}}}{{A_{{n_{\text{R}}}}^e}}} {e^{ - jk\left( {{d_{{n_{\text{T}}},m,n}} + {d_{{n_{\text{R}}},m,n}}} \right)}} \hfill \\
  {\kern 1pt} {\kern 1pt} {\kern 1pt} {\kern 1pt} {\kern 1pt} {\kern 1pt} {\kern 1pt} {\kern 1pt} {\kern 1pt} {\kern 1pt} {\kern 1pt} {\kern 1pt} {\kern 1pt} {\kern 1pt} {\kern 1pt} {\kern 1pt} {\kern 1pt} {\kern 1pt} {\kern 1pt} {\kern 1pt}  = \frac{{\sqrt {{g_{{n_{\text{T}}}}}{g_{m,n}}{p_{{n_{\text{T}}}}}{d_y}{d_z}} 2{\kappa _0}{\Gamma _{m,n}}}}{{4\pi {d_{{n_{\text{T}}},m,n}}{d_{{n_{\text{R}}},m,n}}}}{e^{ - jk\left( {{d_{{n_{\text{T}}},m,n}} + {d_{{n_{\text{R}}},m,n}}} \right)}} \hfill \\
  {\kern 1pt} {\kern 1pt} {\kern 1pt} {\kern 1pt} {\kern 1pt} {\kern 1pt} {\kern 1pt} {\kern 1pt} {\kern 1pt} {\kern 1pt} {\kern 1pt} {\kern 1pt} {\kern 1pt} {\kern 1pt} {\kern 1pt} {\kern 1pt} {\kern 1pt} {\kern 1pt} {\kern 1pt} {\kern 1pt}  \times {F_{{n_{\text{T}}}}}\left( {\theta _{m,n}^{\text{T}},\phi _{m,n}^{\text{T}}} \right){F_{m,n}}\left( {\theta _{m,n}^{{\text{in}}},\phi _{m,n}^{{\text{in}}}} \right) \hfill \\
  {\kern 1pt} {\kern 1pt} {\kern 1pt} {\kern 1pt} {\kern 1pt} {\kern 1pt} {\kern 1pt} {\kern 1pt} {\kern 1pt} {\kern 1pt} {\kern 1pt} {\kern 1pt} {\kern 1pt} {\kern 1pt} {\kern 1pt} {\kern 1pt} {\kern 1pt} {\kern 1pt} {\kern 1pt} {\kern 1pt}  \times {F_{m,n}}\left( {\theta _{m,n}^{{\text{out}}},\phi _{m,n}^{{\text{out}}}} \right){F_{{n_{\text{R}}}}}\left( {\theta _{m,n}^{\text{R}},\phi _{m,n}^{\text{R}}} \right) \hfill \\
  {\kern 1pt} {\kern 1pt} {\kern 1pt} {\kern 1pt} {\kern 1pt} {\kern 1pt} {\kern 1pt} {\kern 1pt} {\kern 1pt} {\kern 1pt} {\kern 1pt} {\kern 1pt} {\kern 1pt} {\kern 1pt} {\kern 1pt} {\kern 1pt} {\kern 1pt} {\kern 1pt} {\kern 1pt}  = \frac{{\beta _{m,n}^{{n_{\text{T}}},{n_{\text{R}}}}{\mathcal{F}}_{m,n}^{{n_{\text{T}}},{n_{\text{R}}}}}}{{4\pi {d_{{n_{\text{T}}},m,n}}{d_{{n_{\text{R}}},m,n}}}}{e^{ - jk\left( {{d_{{n_{\text{T}}},m,n}} + {d_{{n_{\text{R}}},m,n}} - {\varphi _{m,n}}} \right)}} \hfill \\
\end{gathered}
\end{footnotesize}
\end{equation}
where,
\begin{equation} \label{eq12}
  \beta _{m,n}^{{n_{\text{T}}},{n_{\text{R}}}} = 2{\kappa _0}{A_{m,n}}\sqrt {{g_{{n_{\text{T}}}}}{g_{m,n}}{p_{{n_{\text{T}}}}}{d_y}{d_z}},
\end{equation}
which is related to the design parameters of the antennas, and
\begin{equation} \label{eq13}
\begin{gathered}
  {\mathcal{F}}_{m,n}^{{n_{\text{T}}},{n_{\text{R}}}} = {F_{{n_{\text{T}}}}}\left( {\theta _{m,n}^{\text{T}},\phi _{m,n}^{\text{T}}} \right){F_{m,n}}\left( {\theta _{m,n}^{{\text{in}}},\phi _{m,n}^{{\text{in}}}} \right) \hfill \\
  {\kern 1pt} {\kern 1pt} {\kern 1pt} {\kern 1pt} {\kern 1pt} {\kern 1pt} {\kern 1pt} {\kern 1pt} {\kern 1pt} {\kern 1pt} {\kern 1pt} {\kern 1pt} {\kern 1pt} {\kern 1pt} {\kern 1pt} {\kern 1pt} {\kern 1pt} {\kern 1pt} {\kern 1pt} {\kern 1pt} {\kern 1pt} {\kern 1pt} {\kern 1pt} {\kern 1pt} {\kern 1pt} {\kern 1pt} {\kern 1pt} {\kern 1pt} {\kern 1pt}  \times {F_{m,n}}\left( {\theta _{m,n}^{{\text{out}}},\phi _{m,n}^{{\text{out}}}} \right){F_{{n_{\text{R}}}}}\left( {\theta _{m,n}^{\text{R}},\phi _{m,n}^{\text{R}}} \right), \hfill \\
\end{gathered}
\end{equation}
that indicates the directional gain, which is related to the directional coefficients of the antennas and the relative spatial locations of the transceiver nodes.

According to the principle of vector superposition, the sum electric field at the receiving user ${R_{{n_{\text{R}}}}}$ is the result of the superposition of EM waves forwarded by all units of the meta-surface, which can be expressed as
\begin{equation} \label{eq14}
{E^{{n_{\text{R}}}}} = \sum\limits_{m = 1}^M {\sum\limits_{n = 1}^N {E_{m,n}^{{n_{\text{R}}}}} }.
\end{equation}
Then, by combining (\ref{eq10}), (\ref{eq14}), the signal power transmitted form the node ${T_{{n_{\text{T}}}}}$, forwarded by the whole meta-surface, and received at the user ${R_{{n_{\text{R}}}}}$ can also be calculated as
\begin{equation} \label{eq15}
{p_{{n_{\text{R}}}}} = \frac{{{{\left| {{E^{{n_{\text{R}}}}}} \right|}^2}}}{{2{\kappa _0}}}A_{{n_{\text{R}}}}^e = \frac{{{g_{{n_{\text{R}}}}}{\lambda ^2}{{\left| {{E^{{n_{\text{R}}}}}} \right|}^2}}}{{8\pi {\kappa _0}}}.
\end{equation}

Now, suppose that all the transmitting notes has the same normalized radiation power, i.e., ${p_{{n_{\text{T}}}}} = 1$, then, the channel response from each transmitting note to any receiving user (e.g., form ${T_{{n_{\text{T}}}}}$ to ${R_{{n_{\text{R}}}}}$) can be written as
\begin{equation} \label{eq16}
\begin{footnotesize}
\begin{gathered}
{h^{{n_{\text{T}}},{n_{\text{R}}}}} = \sum\limits_{m = 1}^M {\sum\limits_{n = 1}^N {\frac{{\beta _{m,n}^{{n_{\text{T}}},{n_{\text{R}}}}F_{m,n}^{{n_{\text{T}}},{n_{\text{R}}}}}}{{4\pi {d_{{n_{\text{T}}},m,n}}{d_{{n_{\text{R}}},m,n}}}}{e^{ - jk\left( {{d_{{n_{\text{T}}},m,n}} + {d_{{n_{\text{R}}},m,n}} - {\varphi _{m,n}}} \right)}}} }.
\end{gathered}
\end{footnotesize}
\end{equation}
It can be assumed that ${{\mathbf{H}}_{{\text{R,T}}}} \in {\mathbb{C}^{{N_{\text{R}}} \times {N_{\text{T}}}}}$ denotes the channel matrix between all Tx nodes to all Rx users, according to (\ref{eq16}), we can get
\begin{equation} \label{eq17}
\begin{footnotesize}
\begin{gathered}
{{\mathbf{H}}_{{\text{R,T}}}} = {\left[ {\begin{array}{*{20}{c}}
  {{h^{1,1}}}& \cdots &{{h^{{N_{\text{T}}},1}}} \\
   \vdots & \ddots & \vdots  \\
  {{h^{1,{N_{\text{R}}}}}}& \cdots &{{h^{{N_{\text{T}}},{N_{\text{R}}}}}}
\end{array}} \right]_{{N_{\text{R}}} \times {N_{\text{T}}}}},
\end{gathered}
\end{footnotesize}
\end{equation}
It is apparent from this observation that the traditional wireless channel is predominantly shaped by stochastic environmental factors. Consequently, conventional communication algorithms can only passively adapt to the inherently unpredictable nature of the channel. However, the introduction of meta-surfaces has revolutionized this paradigm by enabling active manipulation of the channel environment to align with the specific conditions and needs of the transceiver. According to (\ref{eq16}), the meta-surface has become part of the wireless channel environment. This active control is achieved by dynamically adjusting the reflection coefficient ${\Gamma _{m,n}}$ of each reconfigurable element, empowering us to exert precise and tailored influence over the channel characteristics.

\section{Low Correlation Multi-Pair Links with Vortex Beams: Simulation to Implementation}
\subsection{Channel Capacity of the Meta-Surface-Aided IoT System with Vortex Beams}
In order to mitigate the correlation between different Tx nodes, traditional precoding methods in the digital domain have been widely used to optimize signal transmission, but there is also a growing interest in exploring analog domain solutions \cite{precoding}. By manipulating the EM radiation patterns of the antennas at each Tx node, it becomes possible to generate distinct beams, thereby reducing the spatial correlation of the propagated waves. As we know, vortex modes, characterized by their unique phase and angular momentum properties, offer a promising avenue to enhance orthogonality and minimize interference between signals transmitted by different nodes \cite{NTT,chen1}. By utilizing distinct vortex modes on different Tx nodes, significant advancements can be achieved in terms of signal isolation and system performance in multi-pair communication scenarios. Assuming that the transmitting signal vector with vortex features can be denoted as
\begin{equation} \label{eq18}
{\mathbf{s}} = {{\mathbf{Q}}_{\text{T}}}{\mathbf{Px}}
\end{equation}
where ${\mathbf{x}}$ denotes the original multi-stream data from different notes with ${\rm E}\left( {{\mathbf{x}}{{\mathbf{x}}^H}} \right) = {\mathbf{I}}$, ${\mathbf{P}} = {\text{diag}}\left( {\left[ {\sqrt {{p_0}} , \ldots ,\sqrt {{p_{\mathcal{L} - 1}}} } \right]} \right)$ is the energy allocation matrix, and $\mathcal{L}$ denotes the dimension of multiplexed data streams, which is also equal to the total number of vortex topology modes, assuming that $\mathcal{L} = {N_{\text{T}}}$. Specifically, ${{\mathbf{Q}}_{\text{T}}}$ denotes the vortex beamforming matrix, which can be expressed as
\begin{subequations} \label{eq19}
\begin{align}
  {\mathbf{Q}}_{\text{T}}^{{\text{OAM}}} &= {\text{diag}}\left( {\left[ {{\text{B}}{{\text{P}}_1}, \ldots {\text{B}}{{\text{P}}_{{n_{\text{T}}}}},{\text{B}}{{\text{P}}_{{N_{\text{T}}}}}} \right]} \right), \hfill \\
  {\text{B}}{{\text{P}}_{{n_{\text{T}}}}} &= F_{{n_{\text{T}}}}^{{\text{OAM}}}\left( {{\theta ^{\text{T}}},{\phi ^{\text{T}}}} \right){e^{ - jl\Delta \phi }}, {\kern 1pt} {\kern 1pt} {\kern 1pt} {\kern 1pt} {\kern 1pt} {\kern 1pt} {\kern 1pt} {\kern 1pt}
\end{align}
\end{subequations}
where ${\text{B}}{{\text{P}}_{{n_{\text{T}}}}}$ denotes the vortex beam pattern of each Tx node, $F_{{n_{\text{T}}}}^{{\text{OAM}}}\left( {{\theta ^{\text{T}}},{\phi ^{\text{T}}}} \right)$ is the normalized radiation pattern of various vortex beams. Different from the radiation pattern ${F_{{n_{\text{T}}}}}\left( {{\theta ^{\text{T}}},{\phi ^{\text{T}}}} \right)$ of normal directional waves, ${\text{B}}{{\text{P}}_{{n_{\text{T}}}}}$ has additional phase diversity within the main-lobe of the radiation beams, i.e., the ${e^{ - jl\Delta \phi }}$, where $l \in \mathcal{L}$ denotes the mode number of each vortex beam, $\Delta \phi  \in \left\{ {0,2\pi } \right\}$ is azimuthal angle in the spherical coordinate \cite{xiong50}. Hence, the low correlation property between vortex beams actually comes from the diversity of phase distribution in the main radiation lobes.

According to (\ref{eq18}), it is assumed that the data generated by different nodes are independent of each other, and the total transmission power of all nodes is constant. Then, it worth noting that we consider the ideal scenario where the transmitting nodes have perfect knowledge of the Channel State Information (CSI) \cite{estimation,yuen3}. As we know, the accurate channel estimation and the design of channel estimation techniques relying on partial CSI constitute specific challenges in meta-surface-aided systems, which have been extensively investigated and discussed in previous studies \cite{yuen-add1}\cite{yuen-add2}. Therefore, in this paper we only theoretically analyze the channel capacity variation with or without vortex beams, and leave the design of a special channel estimation algorithm for a future separate research. Combining with (\ref{eq15}), (\ref{eq16}), and (\ref{eq17}), the transmitted signal reaches the Rx users after being reflected by the meta-surface, and the entire signal transmission process can be written as
\begin{equation} \label{eq21}
{{\mathbf{y}}_{\text{r}}} = {{\mathbf{G}}_{\text{r}}}{\mathbf{Hs}} + {\mathbf{z}} = {{\mathbf{G}}_{\text{r}}}{\mathbf{H}}{{\mathbf{Q}}_{\text{T}}}{\mathbf{Px}} + {\mathbf{z}},
\end{equation}
where, ${{\mathbf{y}}_{\text{r}}}$ is the receiving signal, ${{\mathbf{G}}_{\text{r}}} \in {\mathbb{C}^{{N_{\text{R}}} \times {N_{\text{R}}}}}$ denotes the receiving gain related to the specific setup of each user according to (\ref{eq14}), ${\mathbf{z}} \in {\mathbb{C}^{{N_{\text{R}}} \times 1}}$ is the additive Gaussian noise at the receiver with zero mean and variance ${\delta _z}$.

In fact, in this IoT system, the locations of all Tx nodes and Rx users are relatively fixed with respect to the meta-surface, and together with the a priori LoS assumption, the state of the channel is relatively stable over a long period of time. Therefore, for the sake of fairness of comparison, we temporarily refrain from optimizing and discrete quantization design of the phase distribution on the meta-surface and use the classical water filling principle to analyze the upper bound of the Shannon capacity of the channel. At this point, the difference in channel capacity comes only from correlation differences between the Tx nodes. Let's assume that
\begin{equation} \label{eq22}
\begin{footnotesize}
\begin{gathered}
C = \mathop {\max }\limits_{{\rm{tr}}\left( {\bf{\Omega }} \right) \le {P_{{\rm{total}}}},{\bf{\Omega }} \ge 0} {\log _2}\det \left[ {{{\bf{I}}_{{N_{\rm{R}}}}}{\rm{ + }}\frac{1}{{\delta _z^2}}\left( {{{\bf{H}}_{{\rm{R,T}}}}{{\bf{Q}}_{\rm{T}}}} \right){\bf{\Omega }}{{\left( {{{\bf{H}}_{{\rm{R,T}}}}{{\bf{Q}}_{\rm{T}}}} \right)}^H}} \right],
\end{gathered}
\end{footnotesize}
\end{equation}
in bps/Hz, where ${\bf{D}} = {\bf{Px}}{{\bf{x}}^H}{{\bf{P}}^H}$, $P_{{\text{total}}}$ denotes the total power at all the Tx nodes. Firstly, with the given ${\Gamma _{m,n}}$ on each element, (\ref{eq22}) can be analyzed by the typical Singular Value Decomposition (SVD) algorithm, i.e., ${{\bf{H}}_{{\rm{R,T}}}}{{\bf{Q}}_{\rm{T}}} = {{\bf{U}}_{\rm{R}}}{\bf{\Lambda V}}_{\rm{T}}^H$, where ${\bf{\Lambda }} = {\rm{diag}}\left( {\left[ {{\nu _1},{\nu _2} \ldots ,{v_{{\rm{rank}}\left( {{{\bf{H}}_{{\rm{R,T}}}}{{\bf{Q}}_{\rm{T}}}} \right)}}} \right]} \right)$ denotes the singular values decomposed by ${{\bf{H}}_{{\rm{R,T}}}}{{\bf{Q}}_{\rm{T}}}$, ${\text{rank}}\left(  *  \right)$ calculates the rank of the channel matrix. Then, the power allocation vector ${\mathbf{P}}$ can be obtained by the water-filling strategy, i.e.,
\begin{subequations} \label{eq23}
\begin{align}
{\bf{D}} &= {\bf{V}}{\rm{diag}}\left\{ {{p_1}, \ldots ,{p_{{\rm{rank}}\left( {{{\bf{H}}_{{\rm{R,T}}}}{{\bf{Q}}_{\rm{T}}}} \right)}}} \right\}{{\bf{V}}^H}, \\
{p_k} &= \max \left( {{1 \mathord{\left/
 {\vphantom {1 {{p_0}}}} \right.
 \kern-\nulldelimiterspace} {{p_0}}} - {{{\delta ^2}} \mathord{\left/
 {\vphantom {{{\delta ^2}} {\nu _k^2}}} \right.
 \kern-\nulldelimiterspace} {\nu _k^2}},0} \right).
\end{align}
\end{subequations}
where, $p_0$ satisfies $\sum\nolimits_{k = 1}^{{\rm{rank}}\left( {{{\bf{H}}_{{\rm{R,T}}}}{{\bf{Q}}_{\rm{T}}}} \right)} {{p_k}}  = {P_{{\rm{total}}}}$, which means the power constraint at the transmitters, $k \in \left\{ {1, \ldots ,{\rm{rank}}\left( {{{\bf{H}}_{{\rm{R,T}}}}{{\bf{Q}}_{\rm{T}}}} \right)} \right\}$. Hence, (\ref{eq22}) can be re-written as
\begin{equation} \label{eq24}
C = \sum\limits_{k = 1}^{{\rm{rank}}\left( {{{\bf{H}}_{{\rm{R,T}}}}{{\bf{Q}}_{\rm{T}}}} \right)} {{{\log }_2}\left( {1 + \frac{{{\nu _k}{p_k}}}{{\delta _z^2}}} \right)} {\kern 1pt} {\kern 1pt} {\kern 1pt} {\kern 1pt} {\kern 1pt} {\kern 1pt} ({\rm{bps}}/{\rm{Hz}}),
\end{equation}

It is worth noting that with distinct specific vortex beams generating methods, the channel capacity results calculated above will also be different. In this paper, we just take the typical Bessel beam in (\ref{eq19}) as an example to compare the channel capacity changes of meta-surface-aided normal directional wave or vortex beam transmission systems. In the simulation, it is assumed that ${A_{m,n}} = 1$, ${\varphi _{m,n}} = 0$ for each element on the meta-surface, and the Tx nodes and Rx users are arranged in an equally spaced linear array, separated on both sides of the meta-surface normal. In this way, the variation in channel capacity for the same settings in the simulation can only come from the difference in correlation between the Tx nodes. Other main parameters used in the simulation are listed in Table \ref{tab1}.

\begin{table}[htbp]
\caption{Main simulation parameters.}
\begin{center}
\begin{tabular}{c|c}
\toprule 
 \textbf{Parameters} & \textbf{Values}  \\
\midrule 
 Max distance from Tx to coordinate origin & 1.5 m  \\
 Max distance from coordinate origin to Rx & 3 m  \\
 Number of the Tx-Rx pairs & 2, 3, 4 \\
 Center frequency of the EM waves & 10 GHz \\
 Interval between transmitting nodes & 8 wavelength \\
 Interval between receiving users & 5 wavelength \\
 Interval between meta-surface elements & 0.5 wavelength \\
\midrule
 Vortex modes sets used by Tx nodes & $\left\{ { - 1, + 1} \right\}$ \\
                                   & $\left\{ {+ 1, + 2, +3} \right\}$ \\
                                   & $\left\{ {+ 1, + 2, +3, +4} \right\}$ \\
\bottomrule 
\end{tabular}
\end{center}
\label{tab1}
\end{table}

\begin{figure}[htbp]
\centering
\includegraphics[width=3.2in]{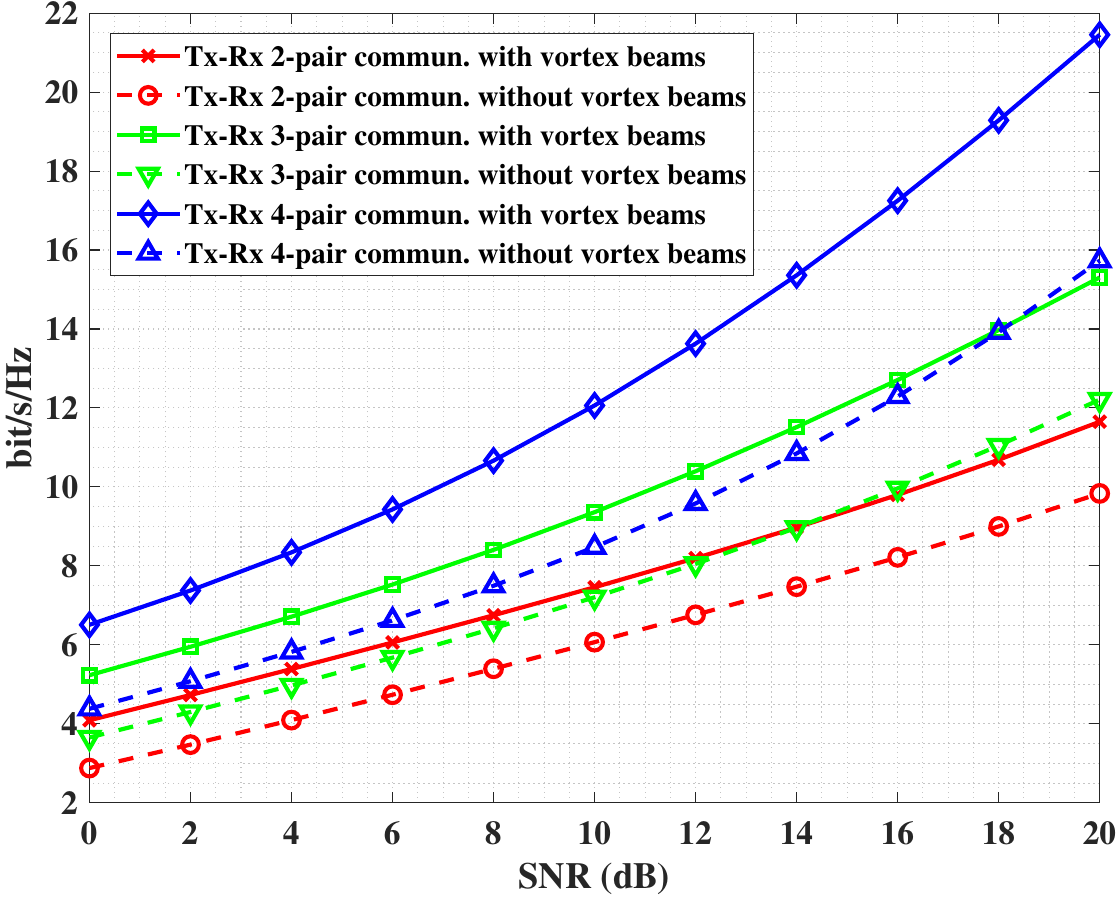}
\caption{Channel capacity variation with SNR for multi-pair.}
\label{fig5}
\end{figure}

Since there is no direct path between the Tx nodes and the Rx users, the received signal completely depends on the forwarding of the meta-surface. Fig. \ref{fig5} gives the calculated results of the channel capacity variation with Signal to Noise Radio (SNR). It can be observed that under the same SNR condition, if all Tx nodes are replaced with vortex beam antennas, the channel capacity will be significantly improved compared to the conventional directional wave antennas. Furthermore, it is worth noting that the capacity of the systems, as depicted by the different curves, exhibits growth with an increasing number of multi-pair links. However, the system utilizing vortex waves demonstrates a significantly faster growth in channel capacity compared to the system employing normal directional waves. This disparity can be attributed to the inherent characteristics of conventional plane wave LoS channels, which exhibit strong correlation under specific transmit power conditions and do not fully exploit the potential of multiplexed communications with multiple data streams. As a result, these channels become a bottleneck for further capacity expansion. In contrast, vortex beams introduce additional variability to the LoS channels through wavefront phase differences among their distinct modes. This leads to a reduction in correlation between different transmission channels, thereby opening up new possibilities for multiplexed multi-stream transmission and enabling substantial enhancements in channel capacity.

\begin{figure}[htbp]
\centering
\includegraphics[width=3.5in]{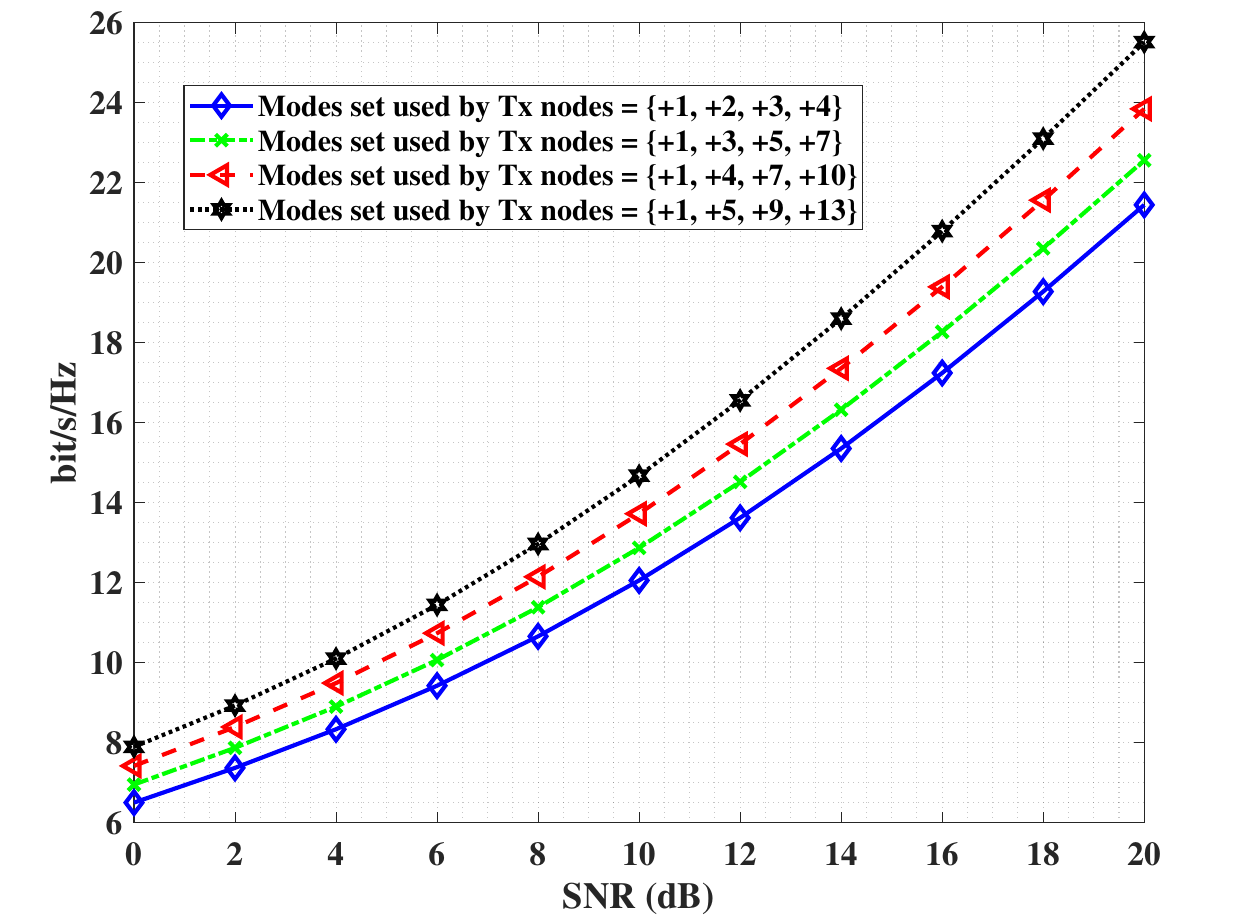}
\caption{Comparison of different vortex mode sets utilized on the Tx nodes.}
\label{fig6}
\end{figure}

As we know, there are few other scattering paths between the Tx nodes and the meta-surface under LoS conditions. The multi-stream data multiplexing transmission capability can be quantified by the channel condition number as
\begin{equation} \label{eq24_1}
{\eta _{{\text{cond}}}} = {{\nu _{k,{\text{Tx}}}^{\max }} \mathord{\left/
 {\vphantom {{\nu _{k,{\text{Tx}}}^{\max }} {\nu _{k,{\text{Tx}}}^{\min }}}} \right.
 \kern-\nulldelimiterspace} {\nu _{k,{\text{Tx}}}^{\min }}},
\end{equation}
where $k \in \left\{ {1,2, \ldots ,{\text{rank}}\{ {{\mathbf{H}}_{\text{R,T}}}{{\mathbf{Q}}_{\text{T}}}\} } \right\}$, ${\nu _{k,{\text{Tx}}}^{\max }}$ and ${\nu _{k,{\text{Tx}}}^{\min }}$ respectively represent the maximum and minimum singular values of the channel matrix after SVD decomposition. It is true that the orthogonality between multiple modes of the vortex beam is being broken in the non-ideal case, e.g., non-alignment, phase deviation, etc., but the low correlation property between the vortex antennas still holds \cite{TVT}.
Fig. \ref{fig6} shows the variation of the total channel capacity of the whole system with SNR, and the different curves indicate that the Tx nodes employ vortex beams with different mode interval. Assume that there are $4$ Tx nodes and $4$ Rx nodes independently. Then curve with mode interval of $1$ means that the $4$ transmitting nodes adopt the set of vortex modes $\left\{ {+ 1, + 2, +3, +4} \right\}$, and so on. Obviously, it can be observed that with a larger mode interval, the channel capacity increases while keeping the distance, emission power, and other simulation settings constant. This observation reinforces the notion that by enhancing the diversity among Tx nodes, the correlation between LoS channels can be mitigated, resulting in a shift in the distribution of channel singular values. This shift is the fundamental mechanism that enables capacity improvement under the same configuration.

It's worth noting that with prior knowledge of the CSI, different optimization objectives can be set to achieve the optimal performance of the meta-surface-aided system under finite constraints. In \cite{active}, researchers explored the joint optimization of active beamforming at the base station and passive beamforming with the meta-surface, aiming to make the most of available CSI. Addressing the challenge of maximizing energy efficiency was the primary goal in \cite{energyef}, where researchers employed a gradient descent approach and sequential fractional programming for this purpose. Moreover, the capacity maximization of a single-user MIMO system with the support of a RIS was the focus of research in \cite{singleuser}, which introduced an alternating optimization algorithm to find an optimal solution. Prior studies have offered valuable insights. In IoT application scenarios, the mathematical modeling of optimization problems presents new characteristics. For instance, in meta-surface-aided multi-pair communications, the optimization objectives can vary depending on specific requirements. They can be either to maximize the overall system rate or to maximize the transmission rate of the smallest node while adhering to interference suppression principles. To better match practical systems, the constraint conditions should encompass not only magnitude constraints on unit reflection coefficients but also introduce bit quantization limitations. This presents a highly intricate yet profoundly practical mathematical challenge. Additionally, the assumption of LoS channels may provide useful prior information to streamline the problem-solving process. These issues are worth further exploration in future research.

\subsection{Implementation of Vortex Beams for IoT Tx Nodes}
In general, the radiation patterns of vortex beams in the RF band have been verified to be expressed in the form of the first kind Bessel beams \cite{lilong}, i.e.
\begin{equation} \label{eq20}
{\text{B}}{{\text{P}}_{{n_{\text{T}}}}} = \gamma {J_l}\left( {k{a_l}\sin {\theta }} \right){e^{ - jl\Delta \phi }},
\end{equation}
where ${J_l}\left(  *  \right)$ denotes the Bessel function, $\gamma $ is the normalized power factor, $k$ is the wave vector, ${{a_l}}$ denotes the radius of the generator, $l$ is the mode number, and $\theta $ denotes the beam divergency angle. In terms of specific implementations, there are may efficient ways to generate vortex beams, e.g., specific traveling-wave waveguide \cite{xiong}, antenna array \cite{chen2}, parabolic reflector \cite{Liang}. Specifically, for different vortex beams generation methods, the radiation pattern shown in (\ref{eq20}) will also vary somewhat. In other words, (\ref{eq20}) represents only a part of the most basic waveforms, and other multifarious structured beams can also be applied in the transmission scheme proposed in this paper. For example, if the planar spiral OAM beams proposed in \cite{xiong50} are utilized, the (\ref{eq20}) will also be adjusted to its corresponding expression.

\begin{figure}[htbp]
\centering
\includegraphics[width=3.2in]{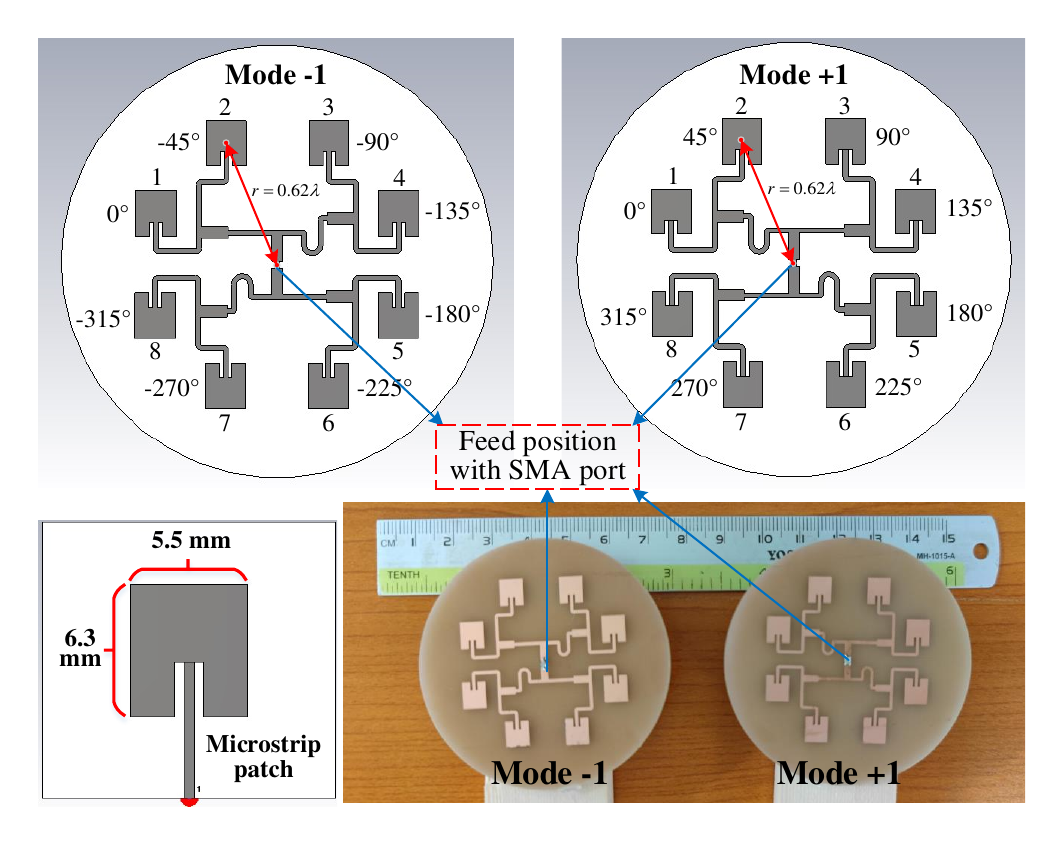}
\caption{Vortex beam generator for each transmitting node (mode $-1, +1$).}
\label{fig3}
\end{figure}

In this work, for simplicity without losing generality, we have designed two different modes of OAM generation antennas at X-band using the mature power-division phase-shifting network. As shown in Fig. \ref{fig3}, 8 microstrip rectangular patch antennas are evenly distributed into a circular shape and connected by a power-division phase-shifting network composed of microstrip lines. By controlling the length of the delay line, the phase difference between distinct patch units is 45 degrees. For the vortex beam array with mode number $+1$, the phase delay increases in clockwise direction. Similarly, for the vortex beam array with mode number $-1$, the phase delay decreases along the counterclockwise direction. The whole power-division phase-shifting network is simulated through the CST Microwave Studio to make it impedance matching. When observing a cross-section taken along the main axis perpendicular to the direction of beam radiation, a distinct circular energy distribution and a spiral spatial phase distribution become evident. Moreover, the vortex beams of various modes ($-1, +1$) produce two completely opposite spatial spiral phase distributions, which represents the orthogonality between the different vortex modes.

\section{Holograph-Inspired Meta-Surfaces for Multi-Pair Communications}
The meta-surface can flexibly steer the incident and outgoing beams by changing the distribution of the reflection coefficients on the whole board. Nevertheless, to ensure that the power of the outgoing signal is not compromised, the meta-surface is always modeled here as a phase-shifting-only surface, which does not block the incident EM waves, thereby maintaining the high efficiency of the transmission signals \cite{song}. Hence, for each Tx node with different location, the phase compensation on the meta-surface can be calculated as
\begin{equation} \label{eq25}
\begin{footnotesize}
\begin{gathered}
\begin{array}{l}
{\varphi _{{n_{\rm{T}}}}}\left( {m,n} \right) = k\sqrt {{{\left( {{x_{m,n}} - {x_{{n_{\rm{T}}}}}} \right)}^2} + {{\left( {{y_{m,n}} - {y_{{n_{\rm{T}}}}}} \right)}^2}{\rm{ + }}{{\left( {{z_{m,n}} - {z_{{n_{\rm{T}}}}}} \right)}^2}} \\
{\kern 1pt} {\kern 1pt} {\kern 1pt} {\kern 1pt} {\kern 1pt} {\kern 1pt} {\kern 1pt} {\kern 1pt} {\kern 1pt} {\kern 1pt} {\kern 1pt} {\kern 1pt} {\kern 1pt} {\kern 1pt} {\kern 1pt} {\kern 1pt} {\kern 1pt} {\kern 1pt} {\kern 1pt} {\kern 1pt} {\kern 1pt} {\kern 1pt} {\kern 1pt} {\kern 1pt} {\kern 1pt} {\kern 1pt} {\kern 1pt} {\kern 1pt} {\kern 1pt} {\kern 1pt} {\kern 1pt} {\kern 1pt} {\kern 1pt} {\kern 1pt} {\kern 1pt} {\kern 1pt} {\kern 1pt} {\kern 1pt} {\kern 1pt} {\kern 1pt} {\kern 1pt} {\kern 1pt} {\kern 1pt} {\kern 1pt} {\kern 1pt} {\kern 1pt} {\kern 1pt} {\rm{ + }}{l_{{n_{\rm{T}}}}}{\tan ^{ - 1}}\left( {\frac{{{z_{m,n}}}}{{{y_{m.n}}}}} \right).
\end{array}
\end{gathered}
\end{footnotesize}
\end{equation}

\begin{figure}[htbp]
\centering
\includegraphics[width=3.5in]{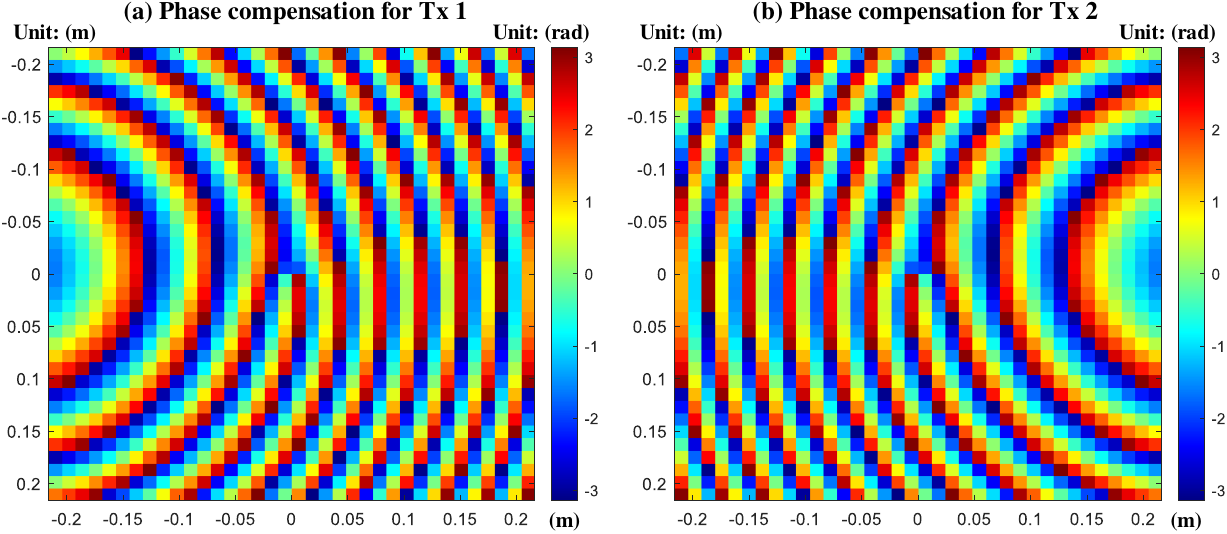}
\caption{Phase compensation on the meta-surface for different Tx notes.}
\label{fig7}
\end{figure}

Figure \ref{fig7} shows the desired phase distributions on the meta-surface for different Tx nodes. Obviously, each node with its unique vortex mode and position requires specific phase compensation, making it challenging to achieve simultaneous phase compensation for all Tx nodes using the same meta-surface. Additionally, the meta-surface faces the daunting challenge of forwarding the compensated signals to different users in order to support multi-pair service requirements in the IoT system, which also poses a significant hardware design challenge for the meta-surface.

Inspired by the holographic technology \cite{holographic,song}, in this paper, we propose a meta-surface design methodology based on the holographic principle. Here, the compensation of the meta-surface can be regarded as a hologram formed by mutual interference between the "reference wave" generated by the Tx node and the "object wave" generated by the Rx user at the meta-surface location. Then, by irradiating the meta-surface with the reference wave, the corresponding object wave is efficiently "reconstructed", thus efficiently establishing a one-to-one mapping communication link from the transmitting node, to the holographic meta-surface, and then to the receiving user.

\subsection{Holographic Recording and Reconstruction}
In accordance with the principles of geometric ray theory, the reference wave irradiate on the $m, n$-th element from the ${n_{\rm{T}}}$-th Tx node can be accurately described as
\begin{equation} \label{eq26}
\begin{footnotesize}
\begin{gathered}
\begin{array}{l}
U_{{\rm{ref}}}^{{n_{\rm{T}}}}\left( {m,n} \right) = \frac{{F_{{n_{\rm{T}}}}^{{\rm{OAM}}}\left( {\theta _{m,n}^{\rm{T}},\phi _{m,n}^{\rm{T}}} \right)}}{{\left| {{{\vec v}_{m,n}} - {{\vec v}_{{n_{\rm{T}}}}}} \right|}}\\
{\kern 1pt} {\kern 1pt} {\kern 1pt} {\kern 1pt} {\kern 1pt} {\kern 1pt} {\kern 1pt} {\kern 1pt} {\kern 1pt} {\kern 1pt} {\kern 1pt} {\kern 1pt} {\kern 1pt} {\kern 1pt} {\kern 1pt} {\kern 1pt} {\kern 1pt} {\kern 1pt} {\kern 1pt} {\kern 1pt} {\kern 1pt} {\kern 1pt} {\kern 1pt} {\kern 1pt} {\kern 1pt} {\kern 1pt} {\kern 1pt} {\kern 1pt} {\kern 1pt} {\kern 1pt} {\kern 1pt} {\kern 1pt} {\kern 1pt} {\kern 1pt} {\kern 1pt} {\kern 1pt} {\kern 1pt} {\kern 1pt} {\kern 1pt} {\kern 1pt} {\kern 1pt} {\kern 1pt} \times {e^{\left( { - jk\left| {{{\vec v}_{m,n}} - {{\vec v}_{{n_{\rm{T}}}}}} \right| - j{l_{{n_{\rm{T}}}}}{{\tan }^{ - 1}}\left( {\frac{{{z_{m,n}}}}{{{y_{m.n}}}}} \right)} \right)}}\\
{\kern 1pt} {\kern 1pt} {\kern 1pt} {\kern 1pt} {\kern 1pt} {\kern 1pt} {\kern 1pt} {\kern 1pt} {\kern 1pt} {\kern 1pt} {\kern 1pt} {\kern 1pt} {\kern 1pt} {\kern 1pt} {\kern 1pt} {\kern 1pt} {\kern 1pt} {\kern 1pt} {\kern 1pt} {\kern 1pt} {\kern 1pt} {\kern 1pt} {\kern 1pt} {\kern 1pt} {\kern 1pt} {\kern 1pt} {\kern 1pt} {\kern 1pt} {\kern 1pt} {\kern 1pt} {\kern 1pt} {\kern 1pt} {\kern 1pt} {\kern 1pt} {\kern 1pt} {\kern 1pt} {\kern 1pt} {\kern 1pt} {\kern 1pt} {\kern 1pt} {\kern 1pt} = {A_{{n_{\rm{T}}}}}{e^{ - jk\Psi _{{\rm{ref}}}^{{n_{\rm{T}}}}\left( {m,n} \right)}}
\end{array}
\end{gathered}
\end{footnotesize}
\end{equation}
where, ${F_{{n_{\rm{T}}}}^{{\rm{OAM}}}\left( {\theta _{m,n}^{\rm{T}},\phi _{m,n}^{\rm{T}}} \right)}$ denotes the normalized radiation pattern for each vortex mode used by the ${n_{\rm{T}}}$-th Tx node, ${{{\vec v}_{m,n}}}$ and ${{{\vec v}_{{n_{\rm{T}}}}}}$ represent the position vectors of the phase center of each element and Tx node, respectively.

Furthermore, assuming that the Rx user is situated in the spherical coordinate system with the center of the meta-surface as the coordinate origin, and its azimuth and pitch angles are denoted by ${\Theta _{{n_{\rm{R}}}}}$ and ${\Phi _{{n_{\rm{R}}}}}$, respectively. At this point, the object wave directed from the meta-surface to the ${n_{\rm{R}}}$-th Rx user can be calculated as
\begin{equation} \label{eq27}
\begin{footnotesize}
\begin{gathered}
U_{{\rm{obj}}}^{{n_{\rm{R}}}}\left( {m,n} \right) = \exp \left( { - jk\left( {\begin{array}{*{20}{c}}
{{y_{m,n}}\sin {\Theta _{{n_{\rm{R}}}}}\cos {\Phi _{{n_{\rm{R}}}}}}\\
{ + {z_{m,n}}\sin {\Theta _{{n_{\rm{R}}}}}\sin {\Phi _{{n_{\rm{R}}}}}}
\end{array}} \right)} \right),
\end{gathered}
\end{footnotesize}
\end{equation}
which is the typical directional beam without vortex wavefront. For each pair of Tx node ${n_{\rm{T}}}$ and Rx user ${n_{\rm{R}}}$, creating a one-to-one mapping channel, the meta-surface records the interferometric holograms of the reference and object waves. This holographic recording process can be represented as,
\begin{equation} \label{eq28}
\begin{footnotesize}
\begin{gathered}
\begin{array}{l}
{O_{{n_{\rm{T}}}\sim{n_{\rm{R}}}}}\left( {m,n} \right) = \frac{{U_{{\rm{obj}}}^{{n_{\rm{R}}}}\left( {m,n} \right)}}{{U_{{\rm{ref}}}^{{n_{\rm{T}}}}\left( {m,n} \right)}}\\
{\kern 1pt} {\kern 1pt} {\kern 1pt} {\kern 1pt} {\kern 1pt} {\kern 1pt} {\kern 1pt} {\kern 1pt} {\kern 1pt} {\kern 1pt} {\kern 1pt} {\kern 1pt} {\kern 1pt} {\kern 1pt} {\kern 1pt} {\kern 1pt} {\kern 1pt} {\kern 1pt} {\kern 1pt} {\kern 1pt} {\kern 1pt} {\kern 1pt} {\kern 1pt} {\kern 1pt} {\kern 1pt} {\kern 1pt} {\kern 1pt} {\kern 1pt} {\kern 1pt} {\kern 1pt} {\kern 1pt} {\kern 1pt} {\kern 1pt} {\kern 1pt} {\kern 1pt} {\kern 1pt} {\kern 1pt} {\kern 1pt} {\kern 1pt} {\kern 1pt} {\kern 1pt} {\kern 1pt} {\kern 1pt} {\kern 1pt} {\kern 1pt} {\kern 1pt} {\kern 1pt} {\kern 1pt} {\kern 1pt} {\kern 1pt} {\kern 1pt} {\kern 1pt} {\kern 1pt} {\kern 1pt} {\kern 1pt} {\kern 1pt} {\kern 1pt} = {a_{{n_{\rm{T}}}\sim{n_{\rm{R}}}}}{e^{\left( {jk\Psi _{{\rm{ref}}}^{{n_{\rm{T}}}}\left( {m,n} \right) - jk\Psi _{{\rm{obj}}}^{{n_{\rm{R}}}}\left( {m,n} \right)} \right)}},
\end{array}
\end{gathered}
\end{footnotesize}
\end{equation}
where, $\Psi _{{\rm{ref}}}^{{n_{\rm{T}}}}\left( {m,n} \right)$ and $\Psi _{{\rm{obj}}}^{{n_{\rm{R}}}}\left( {m,n} \right)$ represent the propagation phases of the reference and object waves on the $m, n$-th element, respectively, while ${a_{_{{n_{\rm{T}}} \sim {n_{\rm{R}}}}}}$ denotes the amplitude compensation term. Specifically, to maximize the energy of the object wave and transmit to the user, each element on the mate-surface maintains the maximum amplitude of the forwarding signal. In other words, it can be constructed as a phase-only holographic surface, where only the phase information is recorded of both the reference and object waves.

Assume that ${N_{\rm{T}}} = {N_{\rm{R}}}$, based on the one-to-one mapping relationship, the total holographic information will be recorded on each unit cell of the meta-surface, i.e.,
\begin{equation} \label{eq29}
\begin{footnotesize}
\begin{gathered}
{O_{{\rm{total}}}}\left( {m,n} \right) = \sum\limits_{{n_{\rm{R}}} = 1}^{{N_{\rm{R}}}} {{O_{{n_{\rm{T}}} \sim {n_{\rm{R}}}}}\left( {m,n} \right)},
\end{gathered}
\end{footnotesize}
\end{equation}
where ${n_{\rm{T}}} \sim {n_{\rm{R}}} \in \left\{ {1 \sim 1,2 \sim 2, \ldots ,{N_{\rm{R}}} \sim {N_{\rm{R}}}} \right\}$, which means the one-to-one mapping multi-pair communication channel. The amplitude normalization progress contributes the phase-only holographic surface, i.e.,
\begin{equation} \label{eq29_1}
\begin{footnotesize}
\begin{gathered}
O_{{\rm{total}}}^{{\rm{norm}}}\left( {m,n} \right) = \frac{{{O_{{\rm{total}}}}\left( {m,n} \right)}}{{\left| {{O_{{\rm{total}}}}\left( {m,n} \right)} \right|}} = \frac{{{\Gamma _{m,n}}}}{{\left| {{\Gamma _{m,n}}} \right|}},
\end{gathered}
\end{footnotesize}
\end{equation}
Hence, according to the well-known Euler's formula, the phase compensation for the $m, n$-th element can be calculated as
\begin{equation} \label{eq30}
\begin{footnotesize}
\begin{gathered}
\begin{array}{l}
{\varphi _{m,n}} = \arg \left[ {O_{{\rm{total}}}^{{\rm{norm}}}\left( {m,n} \right)} \right]\\
{\kern 1pt} {\kern 1pt} {\kern 1pt} {\kern 1pt} {\kern 1pt} {\kern 1pt} {\kern 1pt} {\kern 1pt} {\kern 1pt} {\kern 1pt} {\kern 1pt} {\kern 1pt} {\kern 1pt} {\kern 1pt} {\kern 1pt} {\kern 1pt} {\kern 1pt} {\kern 1pt} {\kern 1pt} {\kern 1pt}  = {\tan ^{ - 1}}\left[ {\frac{{ - \sum\limits_{{n_{\rm{R}}} = 1}^{{N_{\rm{R}}}} {{a_{{n_{\rm{T}}} \sim {n_{\rm{R}}}}}\sin \left( {k\Psi _{{\rm{ref}}}^{{n_{\rm{T}}}} - k\Psi _{{\rm{obj}}}^{{n_{\rm{R}}}}} \right)} }}{{\sum\limits_{{n_{\rm{R}}} = 1}^{{N_{\rm{R}}}} {{a_{{n_{\rm{T}}} \sim {n_{\rm{R}}}}}\cos \left( {k\Psi _{{\rm{ref}}}^{{n_{\rm{T}}}} - k\Psi _{{\rm{obj}}}^{{n_{\rm{R}}}}} \right)} }}} \right]
\end{array}.
\end{gathered}
\end{footnotesize}
\end{equation}

As analyzed above, during the holographic recording process, the phase information of all object and reference waves is simultaneously recorded on the meta-surface. When the meta-surface is excited by the reference wave emitted from one of the nodes, the corresponding object wave is generated in the specified direction, enabling data transmission service to a specific user. It is worth noting that the generation of specific object waves is contingent upon the predetermined position and vortex mode information of the Tx node. In the event that there are changes in the position or vortex mode among the Tx nodes or the Rx user, the meta-surface necessitates reconfiguration, which further underscores significance of individual and adaptable tuning of each unit cell of the meta-surface. It is assumed that the reference wave in (\ref{eq26}) is still utilized to excite the meta-surface, without loss of generality, ${n_{\rm{T}}} \buildrel \Delta \over = 1$, and the reconstruction process that yields the corresponding object wave, i.e., ${n_{\rm{R}}} \buildrel \Delta \over = 1$, can be expressed as follows,
\begin{equation} \label{eq31}
\begin{scriptsize}
\begin{gathered}
\begin{array}{l}
\hat U_{{\rm{obj}}}^1\left( {m,n} \right) = O_{{\rm{total}}}^{{\rm{norm}}}\left( {m,n} \right)U_{{\rm{ref}}}^1\left( {m,n} \right)\\
{\kern 1pt}  = \frac{{{A_{{\rm{ref,1}}}}}}{{\left| {{O_{{\rm{total}}}}\left( {m,n} \right)} \right|}}{e^{ - jk\Psi _{{\rm{ref}}}^1\left( {m,n} \right)}}\sum\limits_{{n_{\rm{R}}} = 2}^{{N_{\rm{R}}}} {{O_{{n_{\rm{R}}}}}\left( {m,n} \right)} \\
{\kern 1pt}  = \frac{{{e^{ - jk\Psi _{{\rm{obj}}}^1\left( {m,n} \right)}}}}{{\left| {{O_{{\rm{total}}}}\left( {m,n} \right)} \right|}} + \frac{{{A_{{\rm{ref,1}}}}}}{{\left| {{O_{{\rm{total}}}}\left( {m,n} \right)} \right|}}\\
{\kern 1pt} {\kern 1pt}  \times \left[ \begin{array}{l}
\frac{1}{{{A_{{\rm{ref,2}}}}}}{e^{ - jk\left( {{d_1} - {d_2}} \right)}}{e^{ - j\left( {{l_1} - {l_2}} \right){{\tan }^{ - 1}}\left( {\frac{{{z_{m,n}}}}{{{y_{m.n}}}}} \right)}}{e^{jk\Psi _{{\rm{obj}}}^2\left( {m,n} \right)}} + \\
\frac{1}{{{A_{{\rm{ref,3}}}}}}{e^{ - jk\left( {{d_1} - {d_3}} \right)}}{e^{ - j\left( {{l_1} - {l_3}} \right){{\tan }^{ - 1}}\left( {\frac{{{z_{m,n}}}}{{{y_{m.n}}}}} \right)}}{e^{jk\Psi _{{\rm{obj}}}^3\left( {m,n} \right)}} +  \\
\ldots
\end{array} \right]
\end{array}.
\end{gathered}
\end{scriptsize}
\end{equation}

It can be seen that ${{{A_{{\rm{obj,1}}}}{e^{ - jk\Psi _{{\rm{obj}}}^1\left( {m,n} \right)}}} \mathord{\left/
{\vphantom {{{A_{{\rm{obj,1}}}}{e^{ - jk\Psi _{{\rm{obj}}}^1\left( {m,n} \right)}}} {\left| {{O_{{\rm{total}}}}\left( {m,n} \right)} \right|}}} \right. \kern-\nulldelimiterspace} {\left| {{O_{{\rm{total}}}}\left( {m,n} \right)} \right|}}$ is the object wave corresponding to the excitation by the reference wave $U_{{\rm{ref}}}^1\left( {m,n} \right)$, which is used to serving the Rx user ${n_{\rm{R}}} \buildrel \Delta \over = 1$. In addition, the meta-surface is irradiated by object waves from specific nodes, which will also generate interference waves in different directions, as shown in the latter part of (\ref{eq31}). However, upon analyzing the decomposition of (\ref{eq31}), it becomes evident that all the interfering waves exhibit vortex phase modulation characteristics. The presence of a central energy hole within the vortex beam itself, along with the inherent beam divergence characteristics, results in minimal energy of the interfering signal reaching other users, thereby having insignificant impact on the demodulation process. The aforementioned remarks will be substantiated through simulations and actual experimental results.

\subsection{Numerical Simulation Analysis}
For clarify, the holographic recording and reconstruction process of the meta-surface can be initially simulated by MATLAB numerical calculations. To maintain simplicity and without loss of generality, we consider a scenario where two independent nodes are located at the transmitting end. These nodes are symmetrically positioned on opposite sides of the normal vector of the meta-surface. Each of these Tx nodes carries a unique communication signal, designated as a 'reference wave', utilizing different modes of vortex beam. When the two reference waves concurrently impinge on the meta-surface, they yield two distinct reflected signals, termed 'object waves', in varying directions. We posit that the two object waves are deflected at angles of ${45^\circ}$ and ${-10^\circ}$ relative to the normal vector of the meta-surface. Correspondingly, the receiving user is situated in the respective directions of these two object waves.

\begin{figure}[htbp]
\centering
\includegraphics[width=3.5in]{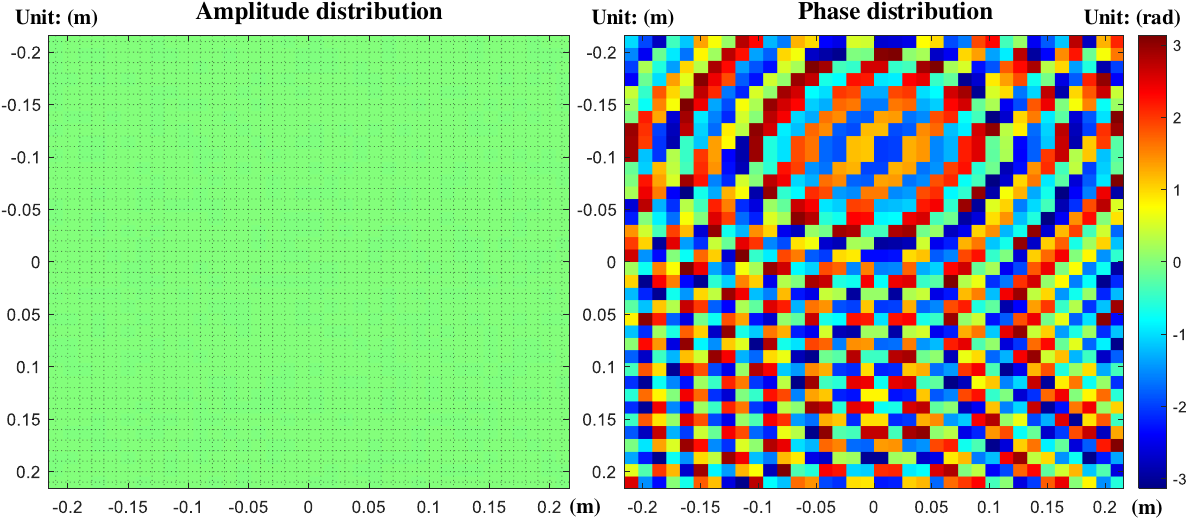}
\caption{Amplitude and phase distribution on the meta-surface.}
\label{fig8}
\end{figure}

In the simulation, the entire meta-surface measures $43.2 \times 43.2$ $\text{cm}$ and consists of $36 \times 36$ individual unit cells. The coordinates of these two Tx nodes are ${{\bf{u}}_{{\rm{T,1}}}} = \left( {0, - 0.25,0.3} \right)$ $\text{m}$ and ${{\bf{u}}_{{\rm{T,1}}}} = \left( {0, 0.25,0.3} \right)$ $\text{m}$, separately. The center frequency is set as 10 $\text{GHz}$. As shown in Fig. \ref{fig8}, after normalization, the amplitudes of all unit cells on the entire meta-surface are equal, while only the phase information is retained and varied.

\begin{figure}[htbp]
\centering
\includegraphics[width=3.5in]{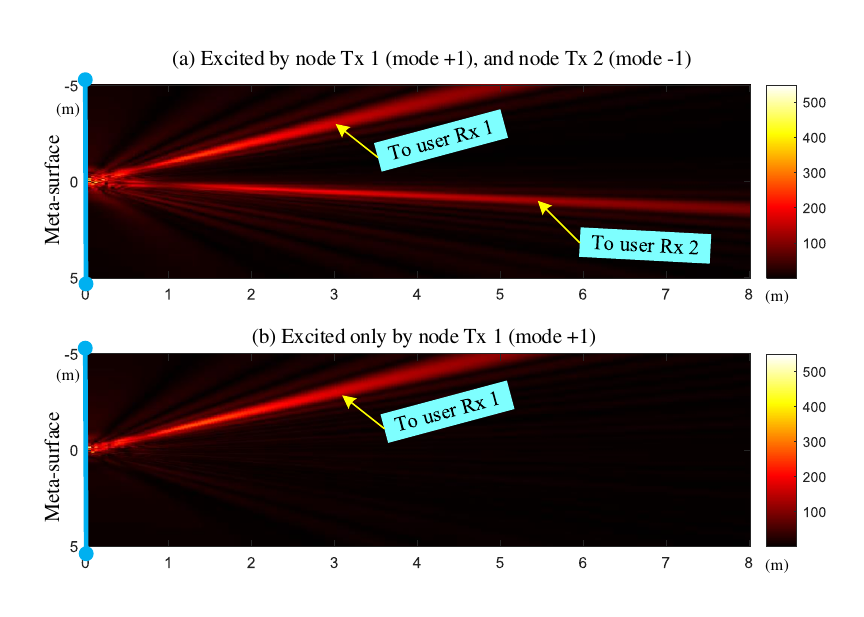}
\caption{Holographic reconstruction progress of the object beams to the users.}
\label{fig9}
\end{figure}

The two Tx nodes generate vortex beams modes of $+1$ and $-1$, respectively, which irradiate the meta-surface simultaneously. After holographic phase compensation, the pointing beam excited by the vortex mode $+1$ is reflected in the direction of ${45^\circ}$, while the pointing beam excited by the vortex mode $-1$ is reflected in the direction of ${-10^\circ}$, thus achieving separation in the spatial angular domain. Fig. \ref{fig9} shows the holographic reconstruction progress of the beams. Obviously, in Fig. \ref{fig9}(a), two reflection beams, i.e. the "object waves", are excited and transmitted to two distinct directions. Then, in Fig. \ref{fig9}(b), only one reflection beam is excited by vortex mode $+1$. It can be seen that there is almost no beam energy in the original ${-10^\circ}$ direction, which indicates that the signal from node Tx 1 is reflected by the holographic meta-surface and produces a pointing beam only in the direction corresponding to the receiving user Rx 1, without interfering with other pairs.

\section{Meta-surface-Aided Multi-pair Communications System Implementation}
\subsection{Unit Cells of the Meta-surface}
\begin{figure}[htbp]
\begin{center}
\noindent
\includegraphics[width=80mm]{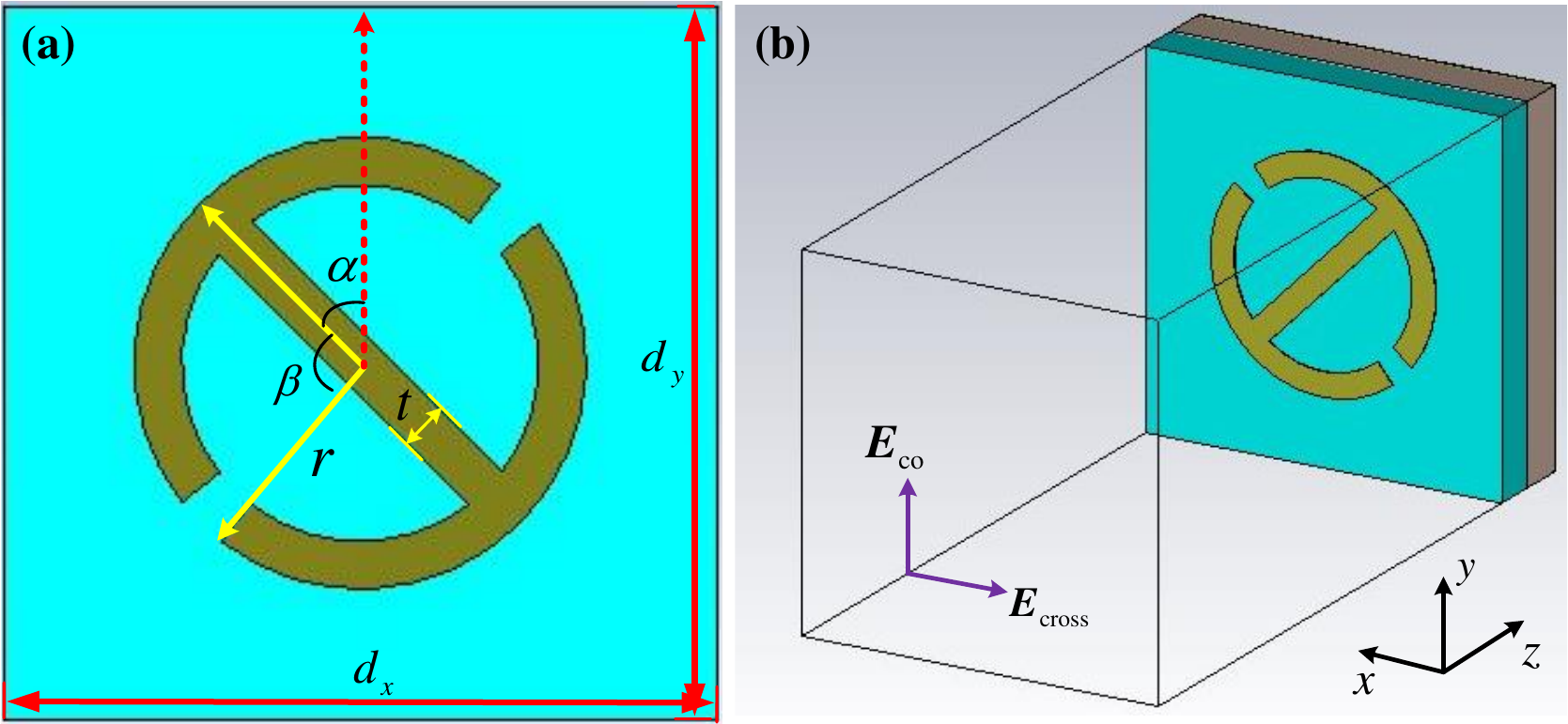} 
\caption{Architecture and simulation configuration of the unit cell. (a) Top view of the I-shaped pattern. (b) View of the simulation setup. $r = 3.8$ mm, $t = 0.8$ mm, and ${d_x} = {d_y} = 12$ mm.}
\label{unit}
\end{center}
\end{figure}

To verify the correctness and effectiveness of the proposed method in IoT scenarios, inspired by the resonant artificial EM structures \cite{cui2}, a kind of modified I-shaped pattern is selected as the fundamental unit for constructing the phase-only holographic meta-surface \cite{Ishape}. As depicted in Fig. \ref{unit}, the unit is composed of two layers: An upper layer with an I-shaped pattern and a lower layer with a copper ground, both situated on dielectric substrates. The orientation angle is denoted by $\alpha $, while the split size is defined by the arc angle $\beta $, and $t = 0.8$ mm is the width of the copper pattern. The dielectric substrates consist of 1.524 mm Rogers 4350B material and 1.6 mm FR-4 board, arranged from top to bottom. This combination of substrates helps to enhance the unit's response bandwidth.

Based on full wave simulations by the CST Microwave Studio, the unit cell can be simulated by implementing periodic boundaries along $x$ and $y$ axes. The reflection parameters are characterized using a Jones matrix \cite{cui2}, which can be expressed as
\begin{equation} \label{eq32}
{\mathbf{J}} = \left[ {\begin{array}{*{20}{c}}
  {{J_{{\text{xx}}}}}&{{J_{{\text{yx}}}}} \\
  {{J_{{\text{xy}}}}}&{{J_{{\text{yy}}}}}
\end{array}} \right].
\end{equation}
Assuming an incident wave polarized in the $x$ direction, the unit cell reflects the wave with cross-polarization in the $y$ direction, represented by $J_{yx}$ in the Jones matrix. Similarly, $J_{xx}$ represents the co-polarization component, satisfying the condition $|J_{xx}|^2 + |J_{yx}|^2 = 1$. By adjusting the orientation angle and split size of the I-shaped pattern, the phase of the cross-polarization reflection wave can be flexibly manipulated, as indicated by $\text{arg}(J_{yx})$. Actually, the arbitrary phase variation brings a large complexity to the system implementation. Hence, many researchers have proposed the concept of $p$-bit-quantized phase distribution, i.e., the continuous phase variation in $2\pi $ period is quantized into ${2^p}$ kinds of discrete phase and implemented with meta-surface cells \cite{bit1}. Extensive studies have demonstrated that the use of 2-bit quantization units employing 2-bit quantization units provides a reliable approximation of the continuous phase distribution of the meta-surface compared to 1-bit and higher-order bit quantization, and a trade-off between quantization phase error and complexity can be achieved \cite{bit1,bit2}.
\begin{figure}[htbp]
\begin{center}
\noindent
\includegraphics[width=88mm]{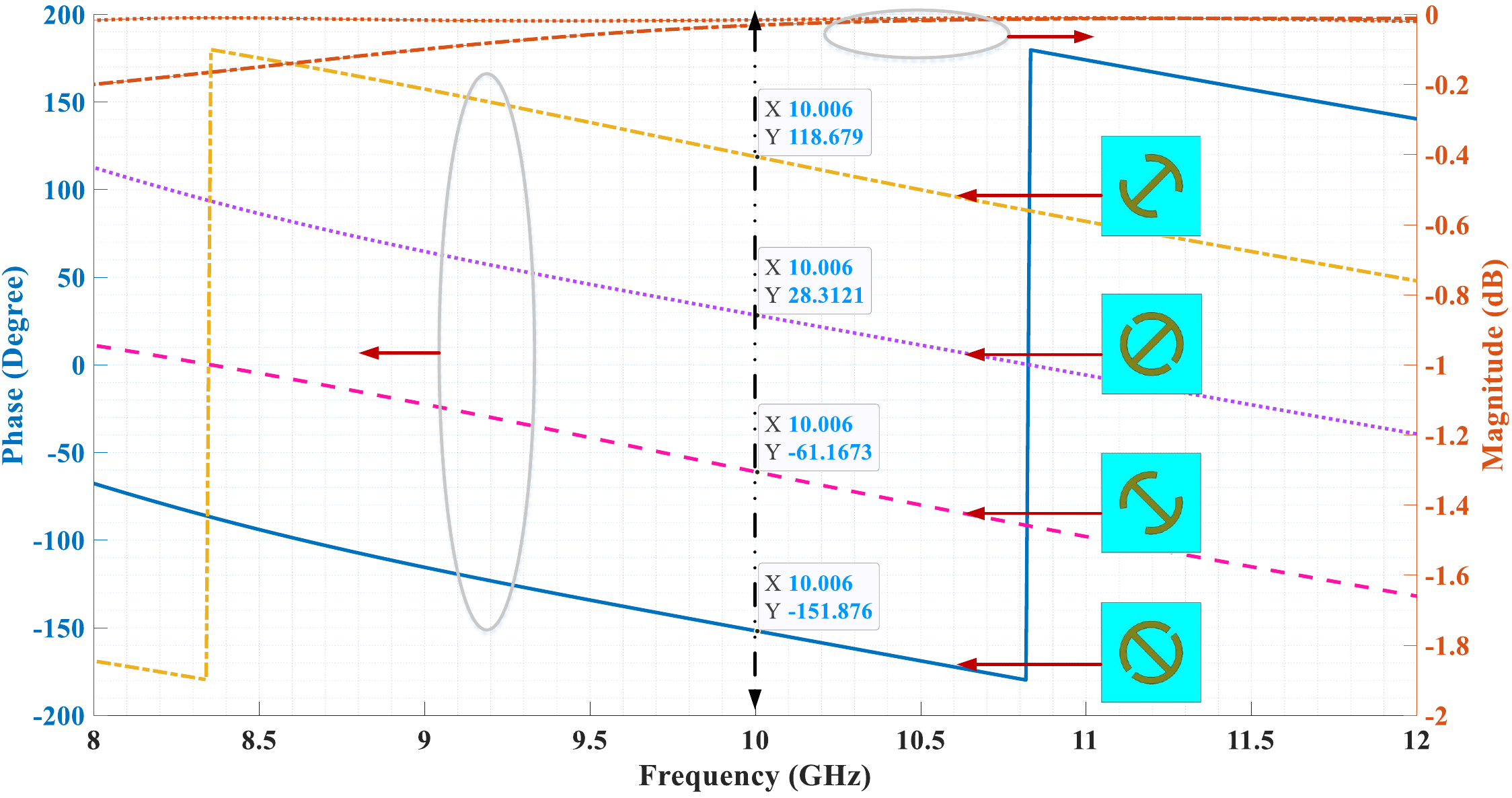} 
\caption{Phase and amplitude responses of the 2-bit phase quantized unit cells with varied $\alpha $ and $\beta $ in frequency domain.}
\label{CST1}
\end{center}
\end{figure}

Therefore, by optimizing the parameters, the reflection phase of the cross-polarization can be adjusted by approximately $90^\circ$ when $\beta = 56^\circ$ or $\beta = 83^\circ$. Additionally, by varying the orientation angle $\alpha$ from $-45^\circ$ to $+45^\circ$, an extra phase difference of $180^\circ$ can be introduced to the reflected signal while maintaining a relatively stable amplitude. Consequently, utilizing these different unit cells enables the realization of phase manipulation capabilities spanning from $0^\circ$ to $360^\circ$, facilitating the construction of a 2-bit quantized reflection meta-surface for beam steering. The full wave simulation is implemented by the CST Microwave Studio. The scan range of the frequency is set at the X-band. As shown in in Fig. \ref{CST1}, the amplitude attenuations of the reflected signal of these four unit cells are less than 0.3 dB within a $40\%$ bandwidth, while the phase differences are stabilized at about $90^\circ$ among them.

\begin{figure*}[!b]
\begin{center}
\noindent
\includegraphics[width=182mm]{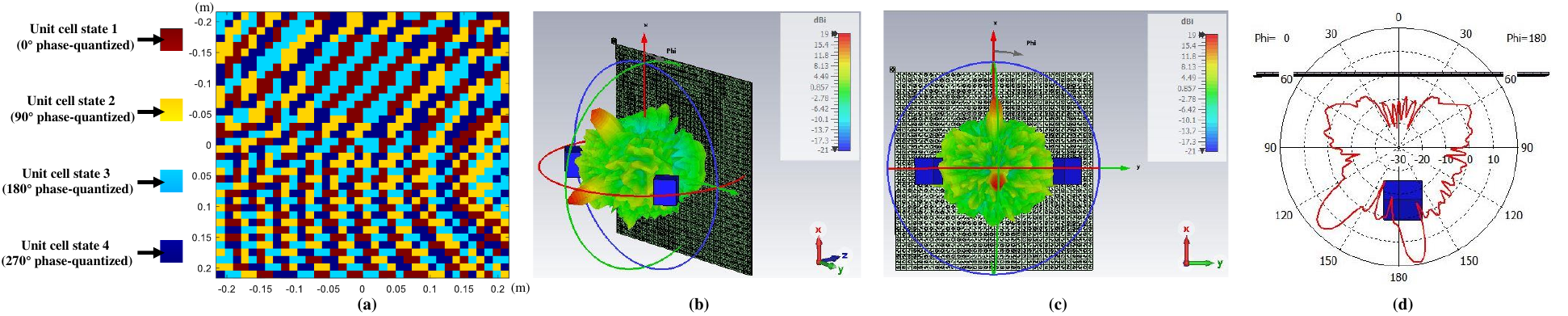} 
\caption{Full waves simulation of the meta-surface-aided two-pair transmission. (a) Holographic phase recordings on the meta-surface. (b)(c) 3D radiation patterns of the reflection beams. (d) 1D radiation patterns in polar coordinate system. }
\label{metasurface}
\end{center}
\end{figure*}

\subsection{Prototyping and Fabrication}
\begin{figure}[htbp]
\begin{center}
\noindent
\includegraphics[width=88mm]{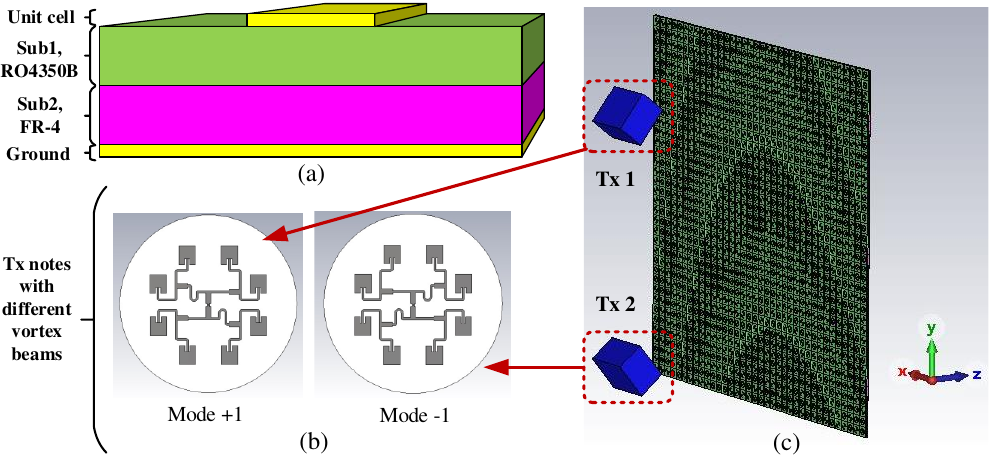} 
\caption{Meta-surface configuration and simulation setup. (a) Layer stack-up. (b) Vortex beam generated by each transmitting note. (c) Simulation scenario in the CST Microwave Studio.}
\label{sub}
\end{center}
\end{figure}

By employing these four unit cell patterns with sequential phase differences of $90^\circ$, we can achieve precise phase manipulation ranging from $0^\circ$ to $360^\circ$. Consequently, this enables the construction of a 2-bit phase-quantized reflection meta-surface, facilitating multi-pair communications in IoT scenarios. As shown in Fig. \ref{sub}(a), the reflective meta-surface structure comprises four layers, with the unit cell pattern printed on a dielectric substrate featuring one ounce thickness of copper. The dielectric substrate is composed of RO4350B and FR-4 materials, while the copper plate is positioned on the reverse side of the substrate. This configuration ensures the wide bandwidth and stable of the reflective meta-surface.

Fig. \ref{sub}(c) shows the simulation scenario in the CST Microwave Studio, the meta-surface consists of $36 \times 36$ individual unit cells, which are distributed according to the principle illustrated in (\ref{eq30}). The phase-quantized mappings are based on Fig. \ref{CST1}. To streamline the calculation time for the Finite Element Method (FEM) in the simulation software, the spatial distribution of the EM field for each individual node is computed separately. These results are then encapsulated into separate data packages and imported into the meta-surface simulation environment. This simulation methodology has been validated and is fully supported by CST Microwave Studio. Adaptive meshing is employed for the FEM calculations, with refined meshing specifically applied at the feeding ports and unit cell patterns. This approach helps to enhance simulation accuracy. Simultaneously, separating each transmission node from the meta-surface reduces computational complexity, resulting in reduced simulation time and reduced dependence on the computer hardware configuration.

According to the study in \cite{OAM}, a circular microstrip antenna array can radiate vortex beam when it is excited by a power-division phase-shifting network. As shown in Fig. \ref{sub}(b), each node employs a circular array to radiate different vortex modes.
In our demonstration, the node Tx 1 uses vortex beam mode $+1$, while the node Tx 2 adopts vortex beam mode $-1$. Each circular array comprises eight identical line-polarized rectangular patch microstrip antennas operating at a central resonant frequency of 10 GHz. The circular array is driven by a multi-stage T-shaped power-division network, allowing for adjustable feed phases. This is achieved by modifying the length of the microstrip lines that feed each patch antenna, resulting in equal intervals of phase increment or decrement. The widths of the feed lines are adjusted to achieve impedance matching. Specifically, for OAM mode $+1$, the feed phases of the eight patch antennas are incrementally increased in a clockwise manner, covering a $360$ degree range. Conversely, for OAM mode $-1$, the feed phases of the patch antennas decrease in a clockwise order, spanning a $-360$ degree range.
The fabrication and full wave simulation results of these vortex beams generators have been illustrated in Fig. \ref{fig3}.

Then, the EM calculation results of each node can be encapsulated into data packets (the blue boxes in Fig. \ref{sub}(c)), which are then imported into the meta-surface simulation environment as signal sources to generate the so called "reference waves". These reference waves are utilized to excite the holographic meta-surface, thereby generating the corresponding "object waves". As depicted in Fig. \ref{metasurface}, the system configuration assumes the presence of two Tx nodes, each generating distinct vortex beam modes that serve as independent reference waves to excite the meta-surface. Following reflection, two separate directional beams are formed, symmetrically distributed at angles of ${45^\circ}$ and ${-10^\circ}$ relative to the normal of the meta-surface, which are indicated in Fig. \ref{metasurface}(b)(c)(d). To ensure experiment reproducibility, we also present the holographic phase distribution results of the meta-surface recordings in Fig. \ref{metasurface}(a).

\begin{figure*}[!t]
\begin{center}
\noindent
\includegraphics[width=180mm]{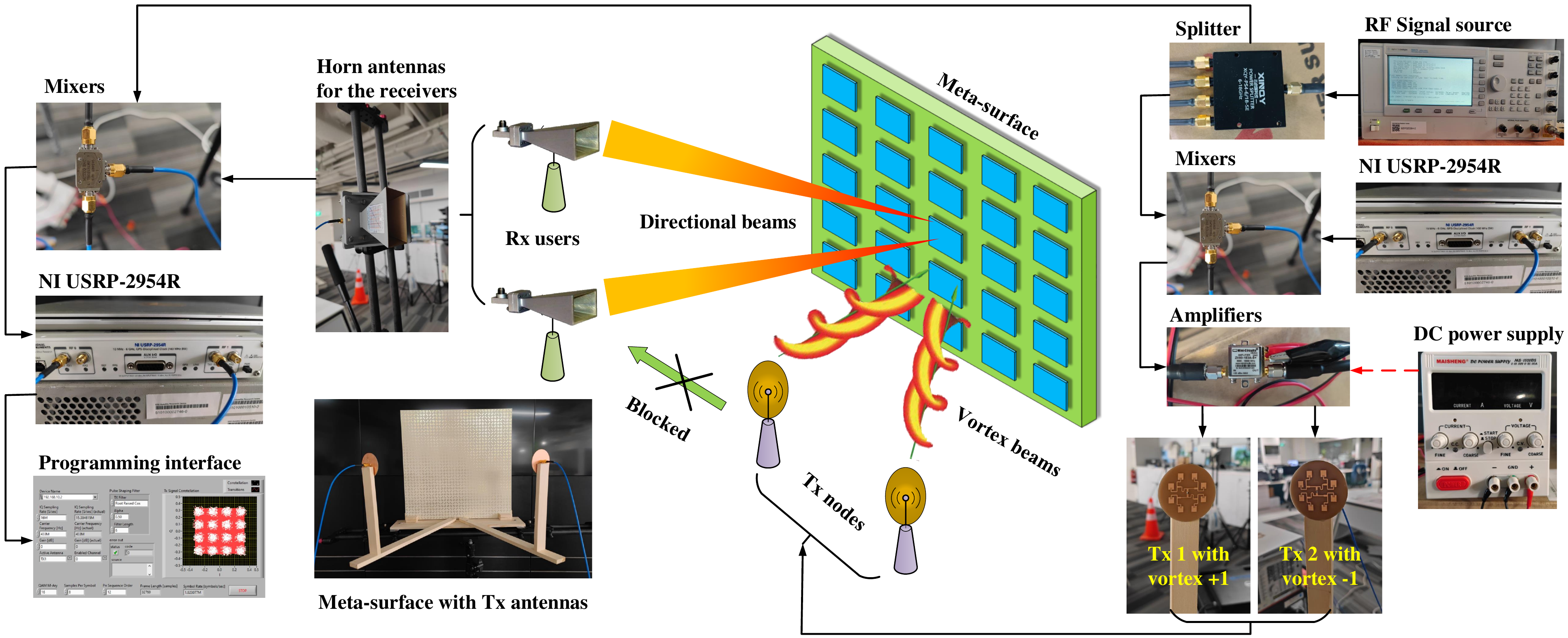} 
\caption{Meta-surface-aided multi-pair IoT Communications with vortex beams, experiment setup and fabrication.}
\label{Fabrication}
\end{center}
\end{figure*}

Furthermore, a real-time communication prototype has been fabricated inside our lab as given in Fig. \ref{Fabrication} and Fig. \ref{Fabrication2}. The two transmitting microstrip antennas are symmetrically positioned on the left and right sides of the normal direction of the meta-surface, with a separation distance of 0.5 m. The vertical distance between the center of each Tx node and the meta-surface is 0.3 m. The meta-surface and the vortex beams emitters are securely mounted on a custom-made fixture at a height of approximately 1.7 m above the ground. The meta-surface itself has dimensions of 43.2 cm by 43.2 cm and is composed of 1296 unit cells arranged evenly in a square configuration. As shown in Fig. \ref{Fabrication}, each vortex beam emitter is connected to a dedicated communication data path through SMA connectors. The communication data, is generated by National Instruments (NI) USRP-2954R and serves as the baseband data stream. It is then fed to the mixer as an intermediate frequency (IF) signal, where it undergoes up-conversion and is connected to the vortex beam emitter. This configuration allows for the transmission of separate communication data stream through different vortex beams. It is worth noted that all mixers receive their reference signals from the same RF signal source (Keysight E8257D). The RF signal is divided into four channels using a power divider, as shown in Fig. \ref{Fabrication}, and subsequently fed to the reference signal inputs of the four mixers (2 Tx notes, 2 Rx users). This ensures that all nodes operate at the same frequency, ensuring precise synchronization and coherence among them.

As mentioned above, the Rx users can utilize conventional RF antennas to capture the directional beams generated through reflection from the meta-surface. In our experiments, two Rx users employed standard X-band horn antennas to receive the reflected beams in their respective directions. The horn antennas were securely mounted on a tripod holder and carefully adjusted for polarization matching, as depicted in Fig. \ref{Fabrication2}. The Rx antennas were positioned at a vertical distance of approximately 3 m from the meta-surface. To determine the most favorable reception areas, the fixed position of the horn antennas was further adjusted using a slide rail, guided by the geometric calculations. This fine-tuning allowed us to identify the two reception regions with the highest energy levels, corresponding to the transmission directions associated with the distinct directional beams.

\subsection{Communication Experiment Results}
\begin{figure}[htbp]
\begin{center}
\noindent
\includegraphics[width=82mm]{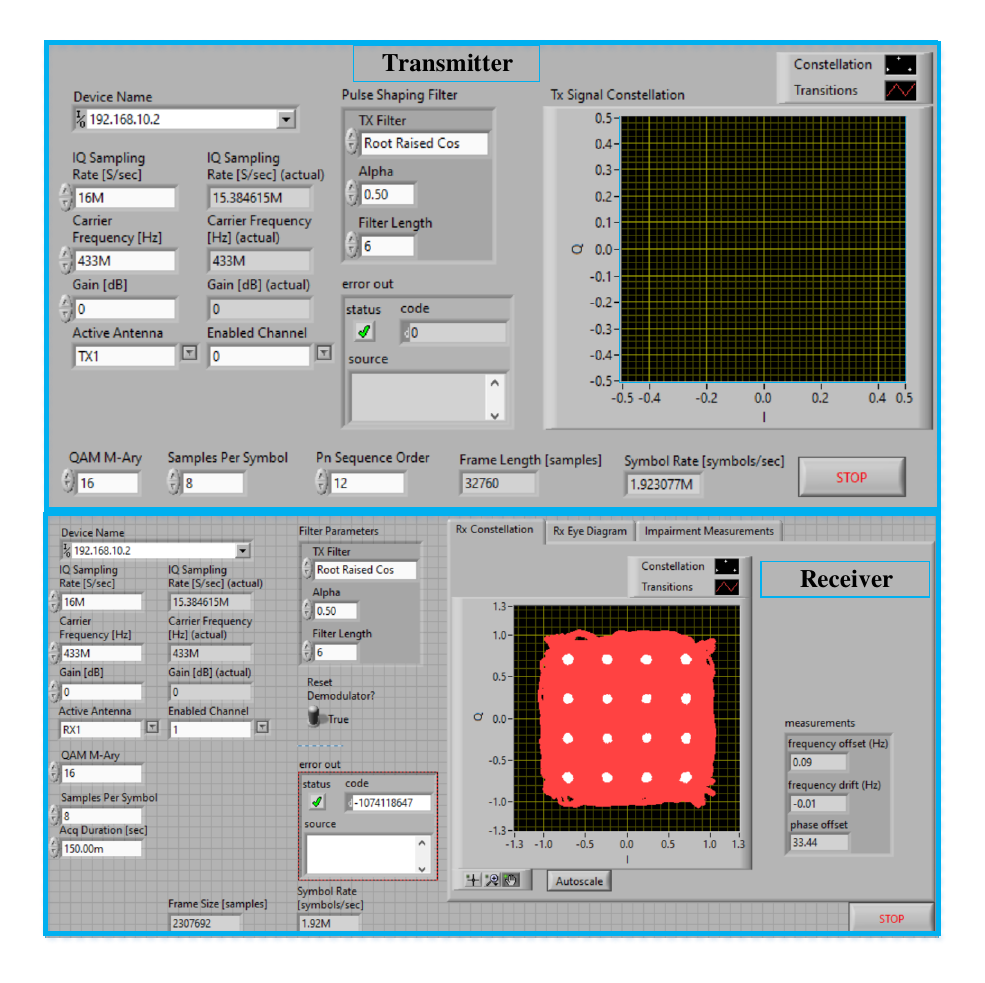} 
\caption{Programming interface and parameter settings for the USRPs.}
\label{interface}
\end{center}
\end{figure}

As we know, the NI USRP-2954R is a software-defined radio device designed to offer flexible and programmable RF capabilities for a wide range of communication applications. It can be configured to transmit and receive RF signals across a broad frequency range, with control and signal processing handled through LabVIEW's graphical programming environment. In the experiment, we used the official LabVIEW communication interface provided by NI, as shown in Fig. \ref{interface}, which allows us to seamlessly control the USRP devices to generate diverse and comprehensive baseband data for conducting rigorous communication tests.

\begin{figure}[htbp]
\begin{center}
\noindent
\includegraphics[width=90mm]{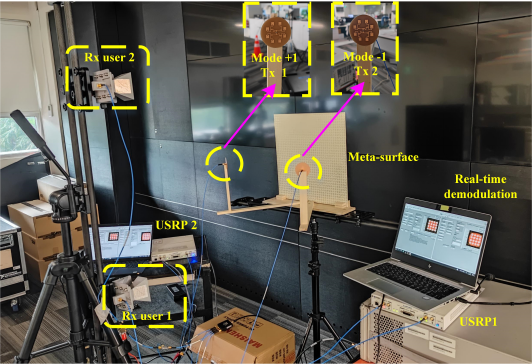} 
\caption{Experiment prototype of the meta-surface-aided 2-pair multi-node communications.}
\label{Fabrication2}
\end{center}
\end{figure}

The real-time communication prototype is shown in Fig. \ref{Fabrication2}. In our experimental setup, we have implemented separate communication data configurations for each Tx node. At the receiver side, we demodulate the two data streams independently and assess the interference between the two communication channels. The baseband data generated by the USRP is fed into the mixer as an IF signal. Following the up-conversion process, the microstrip antenna is excited to emit distinct modes of vortex beams. After reflection by the meta-surface, these vortex beams undergo transformation into general directional beams propagating in different directions. At the receiver end, these beams are coupled and captured by the horn antennas. Subsequently, through the down-conversion operation, which is the inverse of the transmitting process, the received signals are sampled by the USRPs, stored, and demodulated for further analysis.

Some main experimental parameters are summarized in Table \ref{parameter}, providing an overview of the key settings. Both signals are modulated with Quadrature Phase Shift Keying (QPSK) or 16-Quadrature Amplitude Modulation (16-QAM) schemes. The system generates random sequences as data symbols. The frame structure encompasses guide and data segments, while the guide segment facilitates synchronization between the receiver and transmitter, which enables real-time on-line transmission and demodulation, with the demodulation constellation diagram displayed in real-time on the LabVIEW interface depicted in Fig. \ref{interface}.
\begin{table}[htbp]
\caption{Some main experiment parameters.}
\begin{center}
\begin{tabular}{c|c|c}
\toprule
\textbf{Parameter} & \textbf{Value} & \textbf{Dimension} \\
 \midrule
 Central carrier frequency & 10.0 & GHz \\
 Intermediate frequency & 433.0 & MHz \\
 Baseband data bandwidth & 2.0 & MHz \\
 IQ Sampling Rate & 16.0 & MHz \\
 Tx power from the USRP & 0.0 & dBm \\
 Vortex modes & +1, -1 & - \\
 Baseband modulation & QPSK/16-QAM & - \\
 \bottomrule
\end{tabular}
\end{center}
\label{parameter}
\end{table}

\begin{figure}[htbp]
\begin{center}
\noindent
\includegraphics[width=90mm]{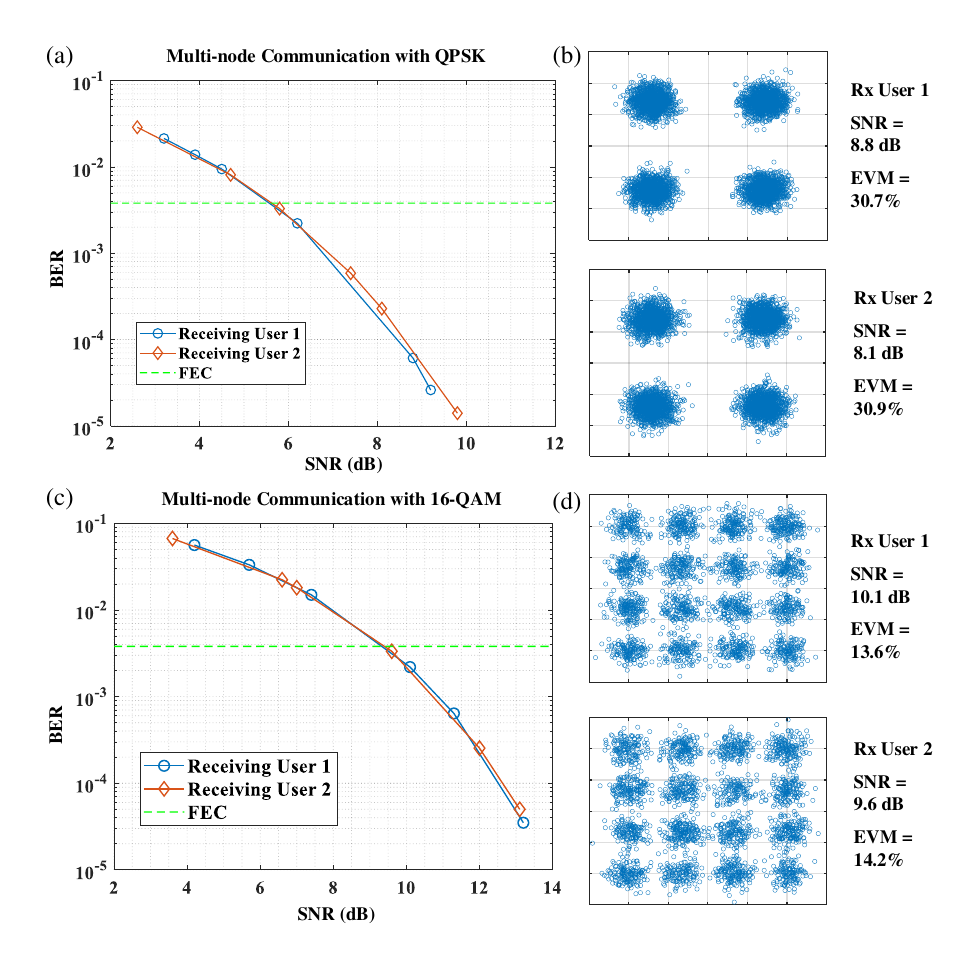} 
\caption{BER performance and demodulation constellation diagram of the Experiment prototype.}
\label{BER}
\end{center}
\end{figure}

By adjusting the transmitting power of USRPs, and together with RF attenuators, we can observe the outcomes at various SNR. The communication data from the two receivers is saved independently on the computer for off-line analysis. Through statistical analysis, we can determine the BER corresponding to different SNR levels, which has been illustrated in Fig. \ref{BER}.

At the Rx users, the receiving power of the two data streams is observed to be nearly equal, indicating a comparable conversion and reflection efficiency of the holographic meta-surface for the two different modes of vortex beams at the same transmitting power. By systematically adjusting the signal power, we conducted a series of experiments generating six distinct sets of data with varying SNRs. In the experiments, two different modulation methods were employed consecutively. Under QPSK modulation, the measured Error Vector Magnitudes (EVM) for the two receiving users are found to be 30.7\% and 30.9\% at SNRs of 8.8 dB and 8.1 dB, respectively, as illustrated in Fig. \ref{BER}(a) and (b). Similarly, for 16-QAM modulation with receiving SNRs of 10.1 dB and 9.6 dB, the corresponding measured EVM are found to be 13.6\% and 14.2\% separately.

Furthermore, the obtained results clearly demonstrate that both curves exhibit the capability to reach the Forward Error Correction (FEC) limit ($3.8 \times 10^{-3}$) \cite{FEC}, thus confirming the feasibility and effectiveness of the communication links. The BER curves for the two receiving users exhibited consistent trends, affirming the system's capability to achieve independent communication between different nodes in IoT scenarios with minimal inter-channel interference. To further evaluate and illustrate the interference between channels, we selectively activated only one transmitting node at a time, simultaneously recording the Received Signal Strength Indicator (RSSI) at both Rx users. The obtained experimental values of RSSI are compiled and presented in Table \ref{interference}.

\begin{table}[htbp]
\caption{Multi-channel isolation measurement}
\begin{center}
\begin{tabular}{c|c|c}
\toprule
\textbf{Transmitting} & \multicolumn{2}{c}{\textbf{RSSI of different users}} \\
\cmidrule{2-3}
\textbf{notes} & Rx user no. 1 & Rx user no. 2 \\
\midrule
Tx no. 1 (vortex +1) & -8.3 dBm & -30.7 dBm \\
\midrule
Tx no. 2 (vortex -1) & -29.8 dBm & -8.6 dBm \\
\midrule
\textbf{Isolations} & 21.5 dB (to Tx no. 2) & 22.1 dB (to Tx no. 1) \\
\bottomrule
\end{tabular}
\label{interference}
\end{center}
\end{table}

Based on the measurement results, it is evident that when node Tx 1 is activated to transmit the mode $+1$ vortex beam, the signal is transmitted through the RF amplifier, then reflected by the meta-surface. The received power measured at user Rx 1 is -8.3 dBm, while the power measured at user Rx 2 is significantly lower at -29.8 dBm, resulting in an isolation of 21.5 dB between the two communication channels. This indicates that the signal transmitted by node Tx 1 has minimal interference on user Rx 2. Similarly, when node Tx 2 is activated to transmit the mode $-1$ vortex beam, the received power measured at user Rx 2 is -8.6 dBm, while the power measured at user Rx 1 is only -30.7 dBm, resulting in an isolation of 22.1 dB between the two channels. This implies that the signal transmitted by node Tx 2 also has minimal impact on the communication quality of user Rx 1. These findings further support the system's ability to facilitate multi-pair multiplexed communication with low interference under LoS channel conditions.

\section{Conclusion and Discussion}
This research introduces a meta-surface-aided multi-pair communications scheme for IoT scenarios. By leveraging holographic-inspired meta-surfaces and vortex beams transmitting antennas, this work addresses the challenge of reducing interference between multi-pair correlated channels. Specifically, the utilization of vortex beams helps minimize correlation under LoS channel conditions, capitalizing on the low correlation property between different topology vortex modes. Moreover, an experimental prototype has been designed for communication links implemented at 10 GHz with QPSK/16-QAM modulation. Two distinct vortex modes have been efficiently excited through microstrip antenna arrays employing the power-division phase-shifting network. The meta-surface consists of a grid of unit cells with specific phase distributions, allowing for precise control of reflected signals. The meta-surface, designed based on the holographic principle, achieves precise 2-bit quantization of the unit cell phase. By effectively separating and conversion multiple vortex beams simultaneously, the meta-surface enables the generation of two independent directional reflection beams. These beams are then utilized to serve distinct users, ensuring efficient and reliable communications with low interference, which has been demonstrated by the experimental results. The receiving BER curves of the two channels have similar trends and can be less than FEC limit ($3.8 \times 10^{-3}$). The isolation between the 2-pair LoS channels can get more than 21 dB.

Due to engineering limitations, the prototype in this paper is constrained to a non-reconfigurable meta-surface and does not include controllable components. In the future, by adopting reconfigurable components such as PIN diodes, variable capacitors, or micro-electro-mechanical systems, these elements provide the advantage of flexible state switching through their connection to voltage regulation circuit networks, which can be controlled by a Field-Programmable Gate Array (FPGA). These developments will be gradually unveiled in our future works. Additionally, in our experiments, we assume knowledge of the relative positions of Tx nodes, the meta-surface, and Rx users. There is extensive research on methods for obtaining CSI. These issues require further investigation and are beyond the scope of our current paper. Finally, the meta-surface compensation method proposed in this paper, based on holographic theory, is not in conflict with optimization methods. In future work, when our prototype gains real-time reconfiguration capabilities, we will delve deeper into designing optimization algorithms based on reconfigurable meta-surfaces, especially in IoT scenarios.
To sum up, the findings highlight the significance of holographic meta-surface design methodology and vortex beams in the context of IoT communications. Moving forward, further exploration and development of meta-surface-aided communication systems are encouraged to unlock their full potential in enabling seamless and robust wireless communication in IoT applications.

\end{document}